\documentclass[aps,prb,twocolumn,preprintnumbers,amsmath,amssymb,superscriptaddress]{revtex4-2}
\usepackage[T1]{fontenc}
\usepackage{xcolor}
\usepackage{textcomp}
\usepackage{bm}
\usepackage{siunitx}
\usepackage{tikz}
\usepackage{yhmath}
\usepackage{multirow}
\usepackage{booktabs}
\usetikzlibrary{patterns,shapes,shadows.blur,calc}
\usepackage{subcaption}
\usepackage{threeparttable}

\definecolor{LinkColor}{rgb}{0.75,0.0,0.2}
\usepackage{hyperref}
\hypersetup{
  colorlinks=true,
  citecolor=LinkColor,
  linkcolor=LinkColor,
  urlcolor=LinkColor,
}

\newcommand{\ket}[1]{\lvert #1\rangle}
\newcommand{\bra}[1]{\langle #1\rvert}
\DeclareMathOperator{\tr}{tr}
\DeclareMathOperator{\Tr}{Tr}

\begin{document}

\title{Noisy Monitored Quantum Circuits}

\author{Shuo Liu}
\affiliation{Department of Physics, Princeton University, Princeton, NJ 08544, USA}
\affiliation{Institute for Advanced Study, Tsinghua University, Beijing 100084, China}

\author{Shao-Kai Jian}
\email{Corresponding Author: sjian@tulane.edu}
\affiliation{Department of Physics and Engineering Physics, Tulane University, New Orleans, Louisiana 70118, USA}

\author{Shi-Xin Zhang}
\email{Corresponding Author: shixinzhang@iphy.ac.cn}
\affiliation{Institute of Physics, Chinese Academy of Sciences, Beijing 100190, China}

\date{\today}
\begin{abstract}
Noisy monitored quantum circuits have emerged as a versatile and unifying framework connecting quantum many-body physics, quantum information, and quantum computation. In this review, we provide a comprehensive overview of recent advances in understanding the dynamics of such circuits, with an emphasis on their entanglement structure, information-protection capabilities, and noise-induced phase transitions. A central theme is the mapping to classical statistical models, which reveals how quantum noise reshapes dominant spin configurations. This framework elucidates universal scaling behaviors, including the characteristic $q^{-1/3}$ entanglement scaling with noise probability $q$ and distinct timescales for information protection. We further highlight a broad range of constructions and applications inspired by noisy monitored circuits, spanning variational quantum algorithms, classical simulation methods, mixed-state phases of matter, and emerging approaches to quantum error mitigation and quantum error correction. These developments collectively establish noisy monitored circuits as a powerful platform for probing and controlling quantum dynamics in realistic, decohering environments.
\end{abstract}
\maketitle

\section{Introduction}
Random quantum circuits (RQCs)~\cite{annurev:/content/journals/10.1146/annurev-conmatphys-031720-030658}, composed of randomly chosen unitary gates, have emerged as a universal tool bridging quantum many-body physics, quantum information, and quantum computing. In the context of quantum many-body physics, RQCs provide a theoretically tractable minimal model for exploring generic aspects of quantum chaos, quantum thermalization, information scrambling, symmetry restoration, and so on~\cite{PhysRevB.92.024301, PhysRevX.7.031016, PhysRevX.8.021014, PhysRevX.8.031058,  PhysRevLett.121.264101,PhysRevX.8.021013, PhysRevX.8.031057, PhysRevB.99.174205, PhysRevB.100.064309,PhysRevLett.123.210601,PhysRevX.10.031066,  PhysRevX.11.011022, PhysRevX.13.041008,PhysRevLett.133.140405, 5d6p-8d1b, PRXQuantum.6.010324, Chen2025subsystem, m3np-p5xj, li2025quantummpembaeffectlongranged, Zhang2025z, Liu_2025, yuQuantumMpembaEffects2025a,aresQuantumMpembaEffects2025a}. RQCs have also been widely applied in quantum sciences, including approximate unitary designs~\cite{PRXQuantum.5.040349, li2024efficientquantumpseudorandomnessconservation, doi:10.1126/science.adv8590,schuster2025strongrandomunitariesfast,cui2025unitarydesignsnearlyoptimal, foxman2025randomunitariesconstantquantum,mao2025randomunitariesconserveenergy, liSUdsymmetricRandomUnitaries2025, zhang2025designsmagicaugmentedcliffordcircuits}, benchmarking for quantum supremacy~\cite{QI_nature19, PhysRevLett.124.080502,PhysRevLett.127.180501,ZHU2022240,morvanPhaseTransitionsRandom2024}, shadow tomography~\cite{huangPredictingManyProperties2020a,elbenRandomizedMeasurementToolbox2023} and variational quantum algorithms~\cite{cerezoVariationalQuantumAlgorithms2021a,RevModPhys.94.015004}.

An important extension of RQCs is the monitored quantum circuits, where unitary dynamics are interspersed with quantum measurements~\cite{PhysRevB.98.205136, PhysRevB.100.134306, PhysRevX.9.031009}. 
In these monitored quantum circuits, the competition between unitary entanglement generation and measurement-induced disentanglement leads to a dynamical phase transition: the steady state entanglement obeys volume law when the measurement probability $p_{m}$ is below the critical probability $p_{m}^{c}$ while it obeys area law when $p_{m}>p_{m}^{c}$~\cite{PhysRevB.98.205136, PhysRevB.100.134306, PhysRevX.9.031009}. This dynamical phase transition is known as measurement-induced phase transitions (MIPTs)~\cite{PhysRevB.98.205136, PhysRevB.100.134306, PhysRevX.9.031009} and has been extensively investigated in monitored quantum circuits~\cite{PhysRevB.99.224307,  PhysRevX.10.041020, PhysRevLett.125.030505, PhysRevB.101.104301, PhysRevB.101.104302, PhysRevB.102.014315,PhysRevB.101.060301, PhysRevB.102.224311, PhysRevLett.125.070606,  PhysRevLett.126.060501, PRXQuantum.2.040319, PhysRevB.103.174309, PhysRevB.103.104306, PhysRevX.12.011045, PhysRevB.105.104306, PhysRevB.106.214316, PhysRevB.106.144311,Zhang_2022,  han2023entanglement, PRXQuantum.5.030311, PhysRevB.109.094209, wang2024drivencriticaldynamicsmeasurementinduced,PhysRevB.110.064311, kelly2023information, kelly2024generalizing,  wang2025relaxationcriticaldynamicsmeasurementinduced, khanna2025randomquantumcircuitstimereversal, xu2025multipartitegreenbergerhornezeilingerentanglementmonitored,Li2025exactaverage,tang2025measurementinducedscramblingemergent}. MIPTs as well as the associated measurement-induced criticality have also been substantially explored in general monitored quantum many-body systems, e.g., monitored SYK model, monitored free fermion and monitored critical ground states~\cite{PhysRevResearch.2.013022, PhysRevLett.127.140601, PhysRevB.103.224210,NonlocalMIPT_Quantum, PhysRevB.107.245132, PhysRevB.102.054302, PhysRevResearch.2.043072, PhysRevB.103.174303, MIPT_SYK_2, Biella_2021, PhysRevB.104.094304, PhysRevX.11.041004, PhysRevLett.128.130605, NonlocalMIPT_Qi,  Zhang2022universal, PhysRevB.106.224305, PhysRevLett.128.010605,PhysRevLett.128.010603,li2023disorderinducedentanglementphasetransitions,PhysRevB.107.094309,  sun2023newcriticalstatesinduced, PhysRevB.108.075151, PhysRevX.13.021026, PhysRevB.107.214203, 10.21468/SciPostPhysCore.7.1.011, PhysRevB.110.035113,  PhysRevB.109.214308,PhysRevLett.133.090401, PhysRevB.110.144305, doggen2023ancilla, PhysRevB.111.184302,Li_2025, ravindranath2023free, PhysRevLett.134.140401,  gopalakrishnan2025monitoredfluctuatinghydrodynamics,PhysRevB.111.014312, foster2025freefermionmeasurementinducedvolumearealaw, gdxd-pw8v, ding2025mixedstatemeasurementinducedphasetransitions,xiao2025topologymonitoredquantumdynamics}. Thanks to the rapid development of quantum simulation hardware, MIPTs have also been demonstrated in several experiments on quantum devices~\cite{noelMeasurementinducedQuantumPhases2022b, kohMeasurementinducedEntanglementPhase2023,hokeMeasurementinducedEntanglementTeleportation2023,PhysRevLett.134.120401,  feng2025postselectionfreeexperimentalobservationmeasurementinduced}.

However, real quantum systems inevitably suffer from environmental decoherence, which fundamentally alters the nature of quantum dynamics from pure-state evolution to mixed-state density matrix evolution. This fact motivates the study of noisy (monitored) quantum circuits. Due to intrinsic randomness, noisy quantum circuits can be mapped to classical statistical models in one higher dimension~\cite{jian2021quantum,PhysRevLett.129.080501,PhysRevB.107.L201113, PhysRevB.108.104310,PhysRevB.108.L060302, PhysRevB.107.014307,PhysRevB.110.064323, PhysRevLett.132.240402}. In this statistical picture, while unitary dynamics promotes permutation symmetry among replicas, the quantum noise plays the role of symmetry breaking field that explicitly favor the identity permutation, thereby competing with entanglement generation. Building on this mapping, the entanglement structure and its universal scaling laws~\cite{PhysRevLett.132.240402,PhysRevB.107.L201113, PhysRevB.108.104310,PhysRevB.108.L060302, PhysRevB.107.014307,PhysRevLett.129.080501,PhysRevLett.134.020403,PhysRevB.110.064323, Guo_2024, he2025measureforgetdynamicsrandom}, the error correction capacity~\cite{ Coding_Vijay, PhysRevLett.132.140401,PhysRevB.110.064323,qian2024coherentinformationphasetransition,nelson2025errorcorrectionphasetransition}, and the classical simulation hardness or nonstabilizerness~\cite{dallas2025nonlocalmagicgenerationinformation,trigueros2025nonstabilizernesserrorresiliencenoisy, PRXQuantum.5.030332,scocco2025risefallnonstabilizernessrandom,zhang2024quantummagicdynamicsrandom,hou2025stabilizerentanglementenhancesmagic,msm2-vmg7,turkeshiMagicSpreadingRandom2025,xiao2026exponentiallyacceleratedsamplingpauli,huang2026fastexactapproachstabilizer} of noisy (monitored) quantum circuits have been extensively explored through both theoretical analysis and numerical simulation.

Beyond the minimal models of brick-wall random circuits, many variations have been explored, motivated by both theoretical considerations and experimental constraints. For instance, imposing conservation laws leads to new dynamical phase transitions, including the charge-sharpening phase transition in symmetric quantum circuits~\cite{PhysRevX.12.041002,PhysRevLett.129.120604, PhysRevB.107.014308,PhysRevB.108.054307, ivakiNoiseResilienceAdaptive2025} with distinct entanglement entropy growth dynamics~\cite{PhysRevLett.122.250602, PhysRevResearch.5.L012031}. Furthermore, introducing measurement feedback allows for the absorbing phase transition, a distinct critical phenomenon that can be detected by linear observables~\cite{PhysRevLett.130.120402,ivakiNoiseResilienceAdaptive2025,PhysRevB.108.L041103, PRXQuantum.4.030318, PhysRevB.109.L020304}. Other model variations investigate the effects of spatiotemporal structure and alternative monitoring protocols, including circuits incorporating temporally correlated randomness~\cite{PhysRevLett.132.240402,  PhysRevB.107.L220204, PhysRevB.108.184204,PhysRevB.110.L140301}, long-range interactions~\cite{PRXQuantum.2.010352,PhysRevLett.128.010604,PhysRevResearch.4.013174,10.21468/SciPostPhysCore.5.2.023, PhysRevResearch.5.L012031}, measurement-only systems~\cite{PhysRevX.11.011030,PRXQuantum.2.030313,lavasaniMeasurementinducedTopologicalEntanglement2021d,PhysRevLett.127.235701,PhysRevResearch.3.023200,PhysRevB.108.115135,PhysRevB.108.094304,PhysRevX.13.041028,PhysRevB.109.024301,Yu_2025,PhysRevB.111.224301,yu2025gaplesssymmetryprotectedtopologicalstates}, weak measurements~\cite{PhysRevB.100.064204, PRXQuantum.5.040313, chenNishimoriTransitionError2025}, and other geometric structures such as quantum trees~\cite{PRXQuantum.4.030333, ravindranath2025entanglementtransitionsnoisyquantum}.

In addition to being a useful tool for studying quantum dynamics, noisy quantum circuits have also inspired developments across a wide range of fields, including variational quantum algorithms~\cite{wangNoiseinducedBarrenPlateaus2021,  Schumann_2024,mele2024noiseinducedshallowcircuitsabsence,sanniaEngineeredDissipationMitigate2024}, classical simulation~\cite{morvanPhaseTransitionsRandom2024, NIPT2_arxiv, PhysRevA.109.042414,PRXQuantum.4.040326,PhysRevLett.133.230403,cichy2025classicalsimulationnoisyquantum, 10.1145/3564246.3585234,PhysRevLett.133.120603,angrisani2025simulatingquantumcircuitsarbitrary, lee2025classicalsimulationnoisyrandom,j1gg-s6zb,GonzalezGarcia2025paulipath,li2025dualrolelowweightpauli,fontana2023classicalsimulationsnoisyvariational}, mixed-state phases of matter~\cite{PhysRevX.14.031044, PhysRevLett.134.070403, ma2025circuitbasedchatacterizationfinitetemperaturequantum, sang2025mixedstatephaseslocalreversibility,PRXQuantum.4.030317,gu2024spontaneoussymmetrybreakingopen,  sala2024spontaneousstrongsymmetrybreaking,PhysRevB.111.064111, PRXQuantum.6.010344, guo2025quantumstrongtoweakspontaneoussymmetry, PRXQuantum.6.010314,PhysRevB.111.L201108, 7p5x-7yqb, liuDiagnosingStrongtoweakSymmetry2025, PhysRevB.111.L060304, ding2026strongtoweakspontaneoussymmetrybreaking}, and quantum error mitigation and correction~\cite{PhysRevLett.119.180509, kimScalableErrorMitigation2023,tran2023localityerrormitigationquantum, qsmz-9kkh, bao2023mixedstatetopologicalordererrorfield, PRXQuantum.5.020343, hlfh-86yz, PhysRevA.111.032402}. Therefore, these models also provide a valuable framework that connects different fields and inspired practical applications.

This review provides a unified overview of recent progress in noisy monitored quantum circuits. 
We begin with basic concepts, including the circuit setup, the relevant observables, and the mapping to classical statistical models. 
We then discuss the entanglement structure of steady states of the noisy monitored quantum circuits, followed by a discussion on the information protection capabilities of such circuits. 
Next, we present the noise-induced phase transitions that arise in these systems, including noise-induced entanglement phase transition, coding transition, and complexity transition. 
Finally, we survey a range of applications of noisy monitored quantum circuits across quantum information science and quantum many-body physics.

\section{Setup of noisy monitored quantum circuits}
\label{sec:Brief introduction to noisy monitored quantum circuits}

We begin by outlining the basic setup of the noisy monitored quantum circuits and the observables used for the investigation of entanglement and information dynamics.

\subsection{Circuit setup}
\label{subsec:Basic setup}

For concreteness, we consider the one-dimensional noisy monitored quantum circuits where the unitary gates are arranged in a brick-wall structure. This 1D system consists of $L$ qudits, each with local Hilbert-space dimension $d$ where $d=2$ corresponds to qubits.
\begin{figure*}[t]
\centering
\includegraphics[width=0.75\textwidth, keepaspectratio]{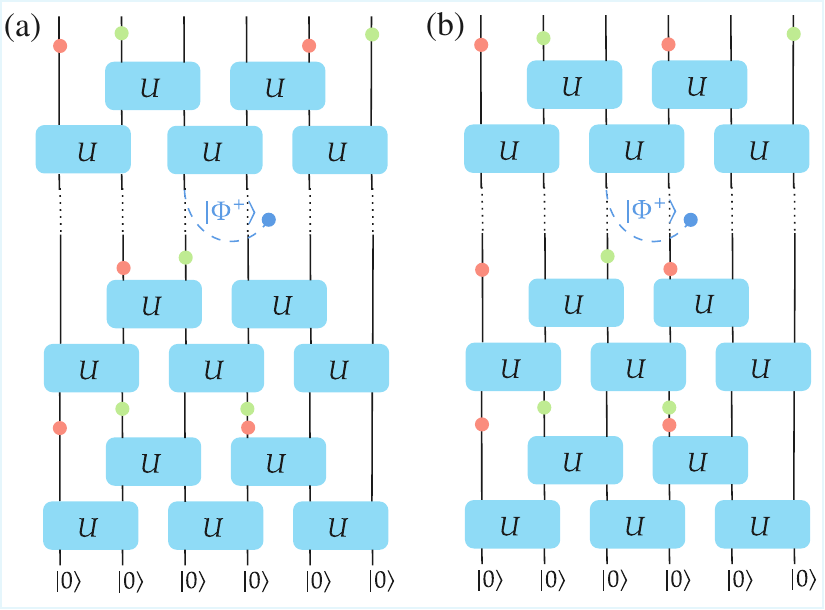}
\caption{\textit{Setup of noisy monitored quantum circuits}. (a) Temporally uncorrelated and (b) correlated quantum noises (red circles). Projective measurements are indicated by green circles. In the steady state encoding scheme, after the system reaches its steady state, a reference qudit (blue circle) is introduced and maximally entangled with the central qudit to form a Bell pair, thereby encoding one-qudit quantum information into the circuit. Reprinted with permission from Ref.~\cite{PhysRevLett.132.240402},
Copyright (2024) by the American Physical Society.}
\label{fig:NoisyQCSetup}
\end{figure*}

To probe entanglement generation, the system is typically initialized in a product state,
\begin{equation}
    \ket{\psi(0)} = \ket{0}^{\otimes L},
\end{equation}
which contains no initial entanglement. Time evolution proceeds in discrete layers, as illustrated in Fig.~\ref{fig:NoisyQCSetup} (a).  
Each layer consists of a brick-wall arrangement of random two-qudit unitary gates acting with periodic boundary conditions.  
Following the unitary layer, both projective measurements and quantum noise channels act independently on every qudit with probabilities $p_{m}$ and $q$, respectively. 
Note that the critical measurement probability reported in Ref.~\cite{PhysRevB.106.214316} is $p_{m}^{c}=0.15995$, which is half of the value in the present setup. This difference arises because their circuit applies a single measurement layer for each (even or odd) unitary layer. 
We denote the total number of time steps by $T$. Note that in this case there is no correlation between quantum noise events at different space-time locations, and the occurrence probability of quantum noise is independent of system size. 
We will also discuss the effects of temporal correlations between noise events shown in Fig.~\ref{fig:NoisyQCSetup} (b) (see Sec.~\ref{sec:Information Protection in Noisy Monitored Circuits}), as well as scenarios in which the occurrence probability becomes system-size dependent (see Sec.~\ref{sec:Noise-induced phase transitions}).

Noisy monitored circuits contain four distinct sources of randomness:
\begin{itemize}
    \item[(1)] independently Haar-random two-qudit unitary gates,
    \item[(2)] random space--time locations of quantum noises,
    \item[(3)] random space--time locations of projective measurements,
    \item[(4)] intrinsically random measurement outcomes dictated by the Born rule.
\end{itemize}
A single realization of all these random elements defines a \emph{trajectory}.  
For any observable of interest, we first compute its value for each trajectory and subsequently average over many such trajectories.  
For nonlinear quantities such as von Neumann entropy, the order of averaging is essential, and we follow the convention described above.

In practice, we choose the evolution time $T$ long enough for entanglement-related observables to saturate, indicating that the system has reached a steady state.  
For monitored circuits without additional quantum noise, this typically requires $T \sim L$~\cite{PhysRevB.98.205136, PhysRevB.100.134306, PhysRevX.9.031009}, although noise can modify this scaling.

\subsection{Entanglement-related observables}
\label{subsec:Entanglement-related observables}

Having introduced the circuit setup, we now describe the observables used to characterize entanglement in noisy monitored quantum circuits.  
Because quantum noise inevitably reduces the purity of the state, the time-evolved density matrix is generically mixed.  
In such settings, conventional entanglement measures based on subsystem von Neumann entropy are no longer reliable to distinguish quantum correlations from classical mixing~\cite{entropy_fail_1, entropy_fail_2}.  
Accordingly, we employ entanglement measures and information-theoretic quantities that remain meaningful for mixed states.

\vspace{2mm}
\noindent\textbf{Logarithmic entanglement negativity.}  
To quantify bipartite entanglement between the left ($A$) and right ($B$) halves of the system, we use the logarithmic entanglement negativity~\cite{Negativity_PhysRevA02, Negativity_PhysRevLett05, Negativity_PhysRevLett12, Negativity_Calabrese, Negativity_PhysRevB19, Negativity_PhysRevLett20, Negativity_PhysRevB20, Negativity_PhysRevLett20_Wu, PRXQuantum.2.030313, Negativity_Shapourian}, defined as
\begin{equation}
    E_{N} = \log \bigl\lVert \rho^{T_{B}} \bigr\rVert_{1},
\end{equation}
where $\rho^{T_{B}}$ is the partial transpose of the density matrix with respect to subsystem $B$, and $\lVert \cdot \rVert_{1}$ is the trace norm.  
Logarithmic entanglement negativity is widely used in the study of monitored circuits~\cite{PhysRevLett.129.080501,PRXQuantum.2.030313} and provides a robust diagnostic of entanglement scaling in the presence of both noise and measurements.

\vspace{2mm}
\noindent\textbf{Mutual information.}  
A complementary diagnostic is the mutual information between $A$ and $B$,
\begin{equation}
    I_{A:B} = S_{A} + S_{B} - S_{AB},
\end{equation}
where $S_{\beta}=-\text{tr}(\rho_{\beta} \ln \rho_{\beta})$ denotes the von Neumann entropy of subsystem $\beta$ with reduced density matrix $\rho_{\beta}$.  
Although mutual information receives contributions from both quantum and classical correlations, it exhibits scaling behavior similar to that of logarithmic entanglement negativity in the investigation of noisy quantum circuits and is particularly natural in the statistical-model description developed later.

\subsection{Information-related observables and encoding schemes}
\label{subsec:Information-protection observables and encoding schemes}
Beyond characterizing entanglement dynamics, it is also essential to quantify the ability of noisy monitored quantum circuits to preserve quantum information.  
In the presence of quantum noise, any encoded information is gradually degraded and eventually leaks into the environment.  
However, how the information-protection timescale depends on the properties of the noise, including its type, occurrence probability, and temporal correlations, is far from obvious and requires a careful operational definition.

To address this question, we consider a standard information-encoding protocol.  
At a designated time $T_{e}$, we introduce a reference qudit $R$, initially prepared in the state $\ket{0}$.  
This reference qudit is then maximally entangled with a chosen system qudit through the creation of a Bell pair.  
Because the circuit is implemented with periodic boundary conditions, the specific spatial location of the system qudit is not important; in all cases, this procedure encodes one qudit of quantum information into the evolving circuit.

We examine two natural encoding schemes:
\begin{itemize}
    \item[(1)] \textit{Steady-state encoding shown in Fig.~\ref{fig:NoisyQCSetup}}, where the Bell pair is inserted at a time $T_{e}>T$, after the circuit has already relaxed to its steady state (see Sec.~\ref{sec:Information Protection in Noisy Monitored Circuits}).
    \item[(2)] \textit{Initial-state encoding shown in Fig.~\ref{fig:Coding}}, where the Bell pair is inserted at $T_{e}=0$ before the circuit evolution begins (see Sec.~\ref{subsec:Noise-induced coding transition}).
\end{itemize}
Following the encoding, the system continues to evolve under the noisy monitored dynamics.  
The central quantity of interest is the characteristic timescale over which the initially encoded information becomes irretrievable due to decoherence.

To quantify how much information remains within the system at time $t$, we evaluate the mutual information between the entire system $AB$ and the reference qudit $R$:
\begin{equation}
    I_{AB:R} = S_{AB} + S_{R} - S_{AB \cup R},
\end{equation}
where $S$ also denotes the von Neumann entropy.  
Immediately after encoding, the formed Bell pair ensures that
\begin{equation}
    I_{AB:R} = 2 \log_d d =2,
\end{equation}
reflecting the full retention of one qudit of quantum information. Note that the logarithm is taken with base $d$ throughout this work. 
In the long-time limit, the decay of this quantity to zero,
\begin{equation}
    I_{AB:R} \rightarrow 0,
\end{equation}
signifies that the encoded information has been completely lost to the environment.  
Tracking the decay of $I_{AB:R}$ therefore provides a direct operational measure of the information-protection capability of noisy monitored quantum circuits.

\section{Mapping to statistical models}
\label{sec:Mapping to statistical models}
In this section, we introduce the mapping between noisy monitored quantum circuits and effective classical statistical models in one higher dimension~\cite{PhysRevLett.129.080501,PhysRevB.107.L201113, PhysRevB.108.104310,PhysRevB.108.L060302, PhysRevB.107.014307,PhysRevB.110.064323, PhysRevLett.132.240402}.  
This mapping provides a unifying framework for understanding entanglement quantities which are encoded in the free energy of the corresponding statistical model.  
For clarity, we first review the mapping for noiseless random unitary circuits and then describe how quantum noise and measurements modify the effective classical model.  
Throughout this subsection, we focus primarily on the von Neumann entropy and mutual information $I_{A:B}$ and $I_{AB:R}$, while the logarithmic entanglement negativity can be treated similarly~\cite{PhysRevB.107.L201113, PhysRevLett.132.240402}.

\subsection{Mapping for noiseless random circuits}
\label{subsec:Mapping for noiseless random circuits}

\begin{figure*}[t]
\centering
\includegraphics[width=0.75\textwidth, keepaspectratio]{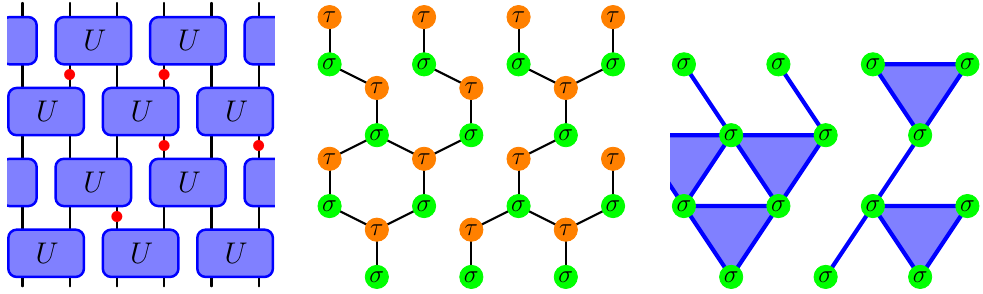}
\caption{\textit{Outline of the mapping to the effective statistical model}.
Starting from the bulk circuit (left), averaging over the unitary gates in Eq.~\eqref{eq:Haaraverage} 
introduces two permutation spins, $\sigma$ and $\tau$, for each two-qubit gate (center). 
Vertical bonds are weighted by the Weingarten function in Eq.~\eqref{eq:Wg}, while diagonal bonds are 
weighted according to Eq.~\eqref{eq:diagonalbond} in the absence of measurements. When a measurement occurs, the 
corresponding diagonal bonds are removed. After integrating out the $\tau$ spins, we obtain a 
model with three-body weights on downward-facing triangles (right) in the absence of measurements; 
these three-body weights reduce to two-body weights when one of the bonds is measured. Reprinted with permission from Ref.~\cite{PhysRevLett.129.080501},
Copyright (2022) by the American Physical Society.
}
\label{fig:OutlineofMapping}
\end{figure*}

Consider a brick-wall circuit composed of Haar-random two-qudit gates, as illustrated in Fig.~\ref{fig:OutlineofMapping} (a).  
After $T$ discrete time steps, the system evolves into
\begin{equation}
    \rho(T)
    =
    \Bigl(\prod_{t=1}^{T} \tilde{U}_{t} \Bigr)
    \rho_0
    \Bigl(\prod_{t=1}^{T} \tilde{U}_{t} \Bigr)^{\!\dagger},
\end{equation}
where $\rho_{0} = \ket{\psi(0)}\bra{\psi(0)}$ is the density matrix of initial state, $\tilde{U}_{t}$ denotes the unitary layer at time step $t$ and is given by
\begin{eqnarray}
    \tilde{U}_{t}= \prod_{i=0}^{\frac{L-2}{2}} U_{t, (2i+2, 2i+3)}  \prod_{i=0}^{\frac{L-2}{2}} U_{t, (2i+1, 2i+2)}.
\end{eqnarray}
We set $L$ to be even without loss of generality, and assume the periodic boundary condition, $U_{t,(L,L+1)} = U_{t,(L,1)}$. 
To perform the Haar average, we vectorize $\rho$ and introduce $r$ replicas (the meaning naturally emerges later):
\begin{equation}
    \ket{\rho(T)}^{\otimes r}
    =
    \prod_{t=1}^{T}
      \bigl[\tilde{U}_{t} \otimes \tilde{U}_{t}^{*} \bigr]^{\otimes r}
    \ket{\rho_0}^{\otimes r}.
\end{equation}
Each random two-qudit unitary gate $U_{t,(i,j)}$ can be averaged independently~\cite{PhysRevB.101.104301, PhysRevB.101.104302, PhysRevX.7.031016, PhysRevLett.129.080501, PhysRevB.99.174205, PhysRevX.12.041002, collins2003moments, collinsIntegrationRespectHaar2006, PhysRevX.8.021014}, giving rise to
\begin{equation}
    \label{eq:Haaraverage}
    \mathbb{E}_{\mathcal U}
    (U \otimes U^{*})^{\otimes r}
    =
    \sum_{\sigma, \tau \in S_{r}}
        \mathrm{Wg}_{d^{2}}^{(r)}(\sigma \tau^{-1})
        \ket{\tau \tau}\!\bra{\sigma \sigma},
\end{equation}
where $\mathrm{Wg}$ is the Weingarten function, $\sigma, \tau$ are permutation spins in the permutation group $S_{r}$ of dimension $r$. 
In the large-$d$ limit, the Weingarten function has an asymptotic expansion~\cite{PhysRevB.99.174205, collinsIntegrationRespectHaar2006}
\begin{equation}
    \label{eq:Wg}
    \mathrm{Wg}_{d^{2}}^{(r)}(\sigma)
    =
    \frac{1}{d^{2r}}
    \left[
        \frac{\mathrm{Moeb}(\sigma)}
             {d^{2|\sigma|}}
        + O(d^{-2|\sigma|-4})
    \right],
\end{equation}
where $|\sigma|$ is the number of transpositions required to construct $\sigma$ from the identity permutation spin $\mathbb{I}$, and $\mathrm{Moeb}(\sigma)$ refers to the M\"obius number of a permutation $\sigma$.

After performing the Haar average, the permutation spins become the effective degrees of freedom, enabling the dynamics of the random quantum circuit to be mapped onto a classical statistical model in one higher dimension.  
The partition function $Z$ of this model is obtained by summing over all permutation-spin configurations, where the weight of each configuration factorizes into contributions from diagonal and vertical bonds, as illustrated in Fig.~\ref{fig:OutlineofMapping}.

The diagonal bond weight corresponds to the overlap between two diagonally adjacent permutation spins,
\begin{equation}
    \label{eq:diagonalbond}
    w_{d}(\sigma,\tau)
    = \langle \sigma | \tau \rangle
    = d^{\, r - |\sigma^{-1}\tau|},
\end{equation}
while the vertical bond weight is determined by the Weingarten function introduced in Eq.~\eqref{eq:Wg}.  
Because the M\"obius function $\mathrm{Moeb}(\sigma)$ appearing in the Weingarten expansion can take negative values~\cite{collinsIntegrationRespectHaar2006}, the resulting bond weights are not manifestly positive.  
To obtain a well-defined classical model with positive statistical weights, we integrate out the intermediate $\tau$ spins, which yields effective three-body weight $W^{0}(\sigma_{1},\sigma_{2};\sigma_{3})$ (see Fig.~\ref{fig:OutlineofMapping}) on downward triangles. In the large-$d$ limit, the leading contributions are
\begin{align}
    W^{0}(\sigma,\sigma;\sigma) &\sim d^{0}, \\
    W^{0}(\sigma',\sigma;\sigma) &\sim d^{-|\sigma'^{-1}\sigma|},
\end{align}
indicating that neighboring permutation spins prefer to align, while domain walls cost an energetic penalty proportional to the permutation distance.  
Thus, the resulting statistical model is effectively ferromagnetic, and its dominant configuration is the minimum-energy arrangement of permutation spins consistent with the imposed boundary conditions.

\subsection{Entanglement entropy from free energy}
\label{subsec:Entanglement entropy from free energy}
We now outline how von Neumann entropy is extracted from the free energy of the statistical model. The von Neumann entropy of a subsystem $\beta$ can be written as
\begin{equation}
    S_{\beta}
    =
    \lim_{n \to 1}
    S_{\beta}^{(n)}
    =
    \lim_{n \to 1}
    \frac{1}{1-n}
    \mathbb{E}_{\mathcal U}
    \log
    \frac{\mathrm{tr}\,\rho_{\beta}^{n}}{(\mathrm{tr}\,\rho)^{n}}.
\end{equation}
Here $S_{\beta}^{(n)}$ is the $n$-th R\'enyi entropy.  
In the replicated Hilbert space, $\mathrm{tr}\,(\rho^{n}_{\beta})$ corresponds to inserting a cyclic permutation $C_{\beta}$ on region $\beta$, while $(\mathrm{tr}\,\rho)^{n}$ corresponds to identity permutations.  
Therefore,
\begin{eqnarray}
    S_{\beta}^{(n)} &=& \frac{1}{1-n} \mathbb{E}_{\mathcal{U}} \log \frac{\tr \rho_{\beta}^{n}}{(\tr \rho)^{n}}  \\ \nonumber
    &=& \frac{1}{1-n} \mathbb{E}_{\mathcal{U}} \log \frac{\Tr((C_{\beta} \otimes I_{\bar{\beta}}) \rho^{\otimes n})}{\Tr ( (I_{\beta} \otimes I_{\bar{\beta}}) \rho^{\otimes n})}  \\ \nonumber
    &=& \frac{1}{1-n} \mathbb{E}_{\mathcal{U}} \log \frac{Z^{(n)}_{\beta}}{Z^{(n)}_{0}},
\end{eqnarray}
where $Z_{\beta}^{(n)}$ and $Z_{0}^{(n)}$ are partition functions with boundary conditions $C_{\beta} \otimes I_{\bar{\beta}}$ and $I_{\beta} \otimes I_{\bar{\beta}}$ respectively.

To move the Haar average inside the logarithm, we introduce an additional replica index $k$~\cite{nishimori2001statistical, kardar2007statistical}:
\begin{align}
    \mathbb{E}_{\mathcal U} \log Z_{\beta}^{(n)}
    &= \lim_{k \to 0} \frac{1}{k} \log Z_{\beta}^{(n,k)}, \\
    \mathbb{E}_{\mathcal U} \log Z_{0}^{(n)}
    &= \lim_{k \to 0} \frac{1}{k} \log Z_{0}^{(n,k)}.
\end{align}
The replicated partition function $Z_{\beta}^{(n,k)}$ corresponds to a classical spin model with fixed top boundary condition $\mathbb{C}_{\beta} = C_{\beta}^{\otimes k} $, while $Z_{0}^{(n,k)}$ corresponds to a uniform identity boundary. Consequently, the entropy reduces to a free-energy difference:
\begin{eqnarray}
    \label{eq:smzz}
    S_{\beta}
    &=&
    \lim_{n \to 1} \lim_{k \to 0}
    \frac{1}{k(1-n)}
    \log \frac{Z_{\beta}^{(n,k)}}{Z_{0}^{(n,k)}}  \\ \nonumber
    &=& \lim_{n \to 1} \lim_{k \to 0}       \frac{
        F_{\beta}^{(n,k)} - F_{0}^{(n,k)}
    }{
        k(n-1)
    }.
\end{eqnarray}
We note that the free energy $F^{(n,k)}$ is proportional to the length of the domain wall, with a unit energy cost of $k(n-1)$. 
As a result, the combination $F^{(n,k)}/[k(n-1)]$ becomes independent of the replica indices $(n,k)$. 
Consequently, the replica limits required to extract the von Neumann entropy in Eq.~\eqref{eq:smzz} can be taken smoothly.

Moreover, because the bottom boundary is free (reflecting the product initial state), the dominant configuration contributing to $F_{0}^{(n,k)}$ is the uniform identity configuration, giving $F_{0}^{(n,k)} = 0$.  
Thus the von Neumann entropy is fully determined by the free energy $F_{\beta}^{(n,k)}$. More importantly, in the large-$d$ limit, partition function $Z_{\beta}^{(n,k)}$ as well as the free energy $F_{\beta}^{(n,k)}$ are determined by the single dominant spin configuration, which greatly simplifies the analytical analysis. This construction establishes the statistical-model framework used to analyze both noiseless and noisy monitored quantum circuits.

To get some intuition from the statistical model, we consider $S_{AB}$. For $\beta = AB$ in the noiseless case, the dominant configuration is uniform with permutation $\mathbb{C}$ throughout, yielding $F_{AB}^{(n,k)} = 0$ and hence $S_{AB} = 0$, consistent with the fact that the quantum state from the output of a noiseless circuit is a pure state.

\subsection{Effects of quantum noise}
\label{subsec:Effects of quantum noise}

We next describe how quantum noise modifies the effective statistical model.  
For concreteness, we illustrate the mapping using the reset quantum channel $\mathcal{R}$, though the resulting structure is independent of the specific choice of local noise channel.

In the replicated description, a reset channel modifies the weight of diagonal bonds connecting diagonally adjacent permutation spins:
\begin{equation}
    \langle \sigma | \mathcal{R} | \tau \rangle
    = d^{\,r - |\tau|}.
\end{equation}
This change propagates into the three-body triangle weights.  
For a downward triangle with identical spins, one finds
\begin{align}
    W^{\mathcal R}(\sigma,\sigma;\sigma)
    &= \sum_{\tau \in S_{r}}
        \mathrm{Wg}_{d^{2}}^{(r)}(\sigma^{-1}\tau)\,
        d^{\,r - |\sigma^{-1}\tau|}\,
        d^{\,r - |\tau|} \nonumber \\
    &\sim d^{-|\sigma|}.
\end{align}
Thus, quantum noise acts as a permutation \emph{symmetry-breaking field}~\cite{jian2021quantum,PhysRevLett.129.080501,PhysRevB.107.L201113, PhysRevB.108.104310,PhysRevB.108.L060302, PhysRevB.107.014307,PhysRevB.110.064323, PhysRevLett.132.240402} that energetically favors the identity permutation $\mathbb{I}$.  
Because noise events occur at random space--time positions, they introduce \emph{quenched disorder} into the classical spin model. Note that quantum noise also relaxes the unitary constraints~\cite{Uaverage_Qi, PRXQuantum.4.010331}, i.e., $W^{0}(\mathbb{C},\mathbb{C};\mathbb{I})=0$ in the absence of quantum noise and thus this configuration is not allowed while $W^{\mathcal{R}}(\mathbb{C},\mathbb{C};\mathbb{I})\sim d^{-\vert \mathbb{C} \vert}$ in the presence of quantum reset channel, and consequently enables additional spin configurations in the presence of noise~\cite{PhysRevLett.132.240402}.

In Sec.~\ref{subsec:Entanglement entropy from free energy}, we have shown that the dominant spin configuration for $\beta = AB$ in the absence of quantum noise is that all spins are $\mathbb{C}$. However, in the presence of quantum noise, this configuration acquires a weight proportional to $d^{-q L T |\mathbb{C}|}$ and is no longer energetically favored. See detailed discussion in Sec.~\ref{sec:Entanglement Structure of Steady States} for the favorable configurations in this case.

\subsection{Effects of projective measurements}
\label{subsec:Effects of projective measurements}
In the presence of projective measurements, the randomness of their space--time locations is likewise treated as quenched disorder. In the effective statistical model, as shown in 
Fig.~\ref{fig:OutlineofMapping}, the corresponding diagonal bonds are removed when projective measurements are applied. Consequently, projective measurements act as a source of random Gaussian potentials and induce fluctuations of the domain wall away from its otherwise optimal trajectory~\cite{PhysRevLett.129.080501}.

Together, these ingredients constitute the statistical‐model description of noisy monitored circuits, enabling both analytical and numerical characterization of entanglement and information dynamics.  
Although the above discussion focuses on extracting entanglement-related observables (such as von Neumann entropy and mutual information) from the effective statistical model, the circuit-to-statistical-model mapping is more general.  
In particular, this framework has also been applied to study the dynamics of magic (nonstabilizerness)~\cite{turkeshiMagicSpreadingRandom2025,zhang2024quantummagicdynamicsrandom} and random circuit sampling~\cite{NIPT2_arxiv, PhysRevA.109.042414} in noisy monitored circuits, 
revealing that the statistical description naturally extends beyond entanglement to other 
nonclassical resources in quantum many-body systems, including magic, fidelity, and cross-entropy benchmarking (XEB)~\cite{boixoCharacterizingQuantumSupremacy2018a,doi:10.1126/science.aao4309,QI_nature19}.

\section{Entanglement Structure of Steady States}
\label{sec:Entanglement Structure of Steady States}

\subsection{KPZ scaling from fluctuating domain walls}
\label{subsec:KPZ scaling from fluctuating domain walls}
\begin{figure*}[t]
\centering
\includegraphics[width=0.75\textwidth, keepaspectratio]{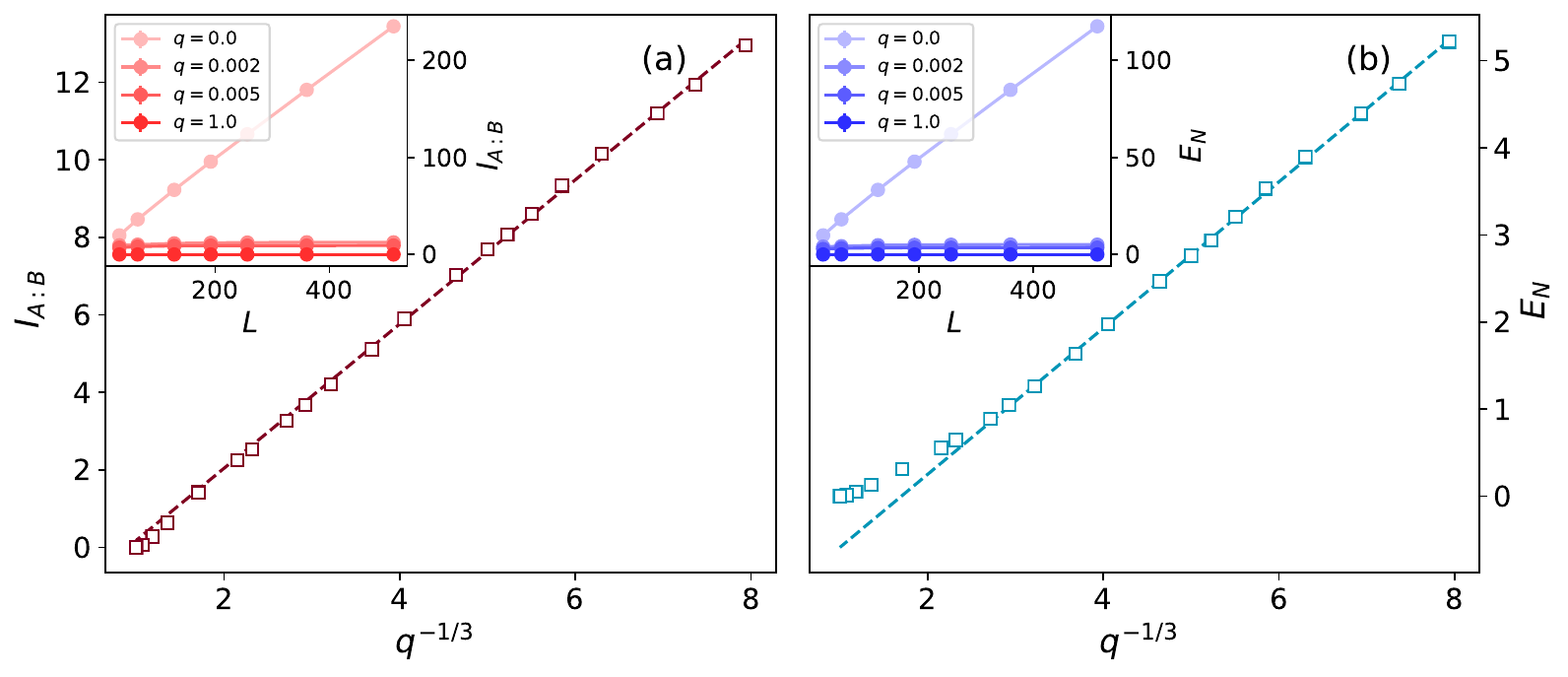}
\caption{\textit{$q^{-1/3}$ entanglement scaling in noisy monitored quantum circuits}.
(a) $I_{A:B}(q)$ and (b) $E_{N}(q)$ of the steady state of noisy monitored quantum circuits. 
The measurement probability is $p_{m} = 0.1<p_{m}^{c}$. 
A characteristic $q^{-1/3}$ scaling emerges for the entanglement in the presence of noise. 
The insets show the dependence of $I_{A:B}$ and $E_{N}$ on the system size $L$ for various noise rates. 
Without noise, the entanglement exhibits a volume-law scaling since $p_{m} < p_{m}^{c}$, whereas any finite noise probability drives the system into an area-law regime. Reprinted with permission from Ref.~\cite{PhysRevB.107.L201113},
Copyright (2023) by the American Physical Society.}
\label{fig:KPZentanglement}
\end{figure*}

In the presence of quantum noise, the entanglement structure of monitored circuits is qualitatively altered. 
Even an arbitrarily small noise rate $q>0$ is sufficient to eliminate the volume-law phase entirely, thereby enforcing an entanglement area law for all values of the projective measurement probability~\cite{PhysRevLett.129.080501,PhysRevB.107.L201113, PhysRevB.108.104310,PhysRevB.108.L060302, PhysRevB.107.014307,PhysRevB.110.064323, PhysRevLett.132.240402}. 
Consequently, the conventional measurement-induced phase transition observed in noiseless circuits disappears. 
Nonetheless, the resulting noise-induced area law is far from trivial: the steady-state entanglement exhibits a universal and highly nontrivial dependence on the noise probability,
\begin{equation}
    I_{A:B}(q) \sim q^{-1/3},
\end{equation}
where $q$ is the quantum noise probability.
Numerical results confirming this scaling are shown in Fig.~\ref{fig:KPZentanglement}.  
Below we outline how this behavior naturally emerges within the classical statistical-model description.

To develop an analytical understanding of the $q^{-1/3}$ scaling, we first examine how quantum noise modifies the dominant spin configurations in the associated statistical model. 
Noise events are treated as quenched disorder in space--time, and the dominant spin configuration is determined independently for each realization. 
As discussed in Sec.~\ref{subsec:Effects of quantum noise}, the fully $\mathbb{C}$ configuration that dominates $F^{(n,k)}_{AB}$ in the absence of noise becomes energetically disfavored once noise is introduced. 
Building on this observation, we now analyze how noise reshapes the dominant spin texture underlying the entanglement scaling.

\begin{figure*}[t]
\centering
\includegraphics[width=0.75\textwidth, keepaspectratio]{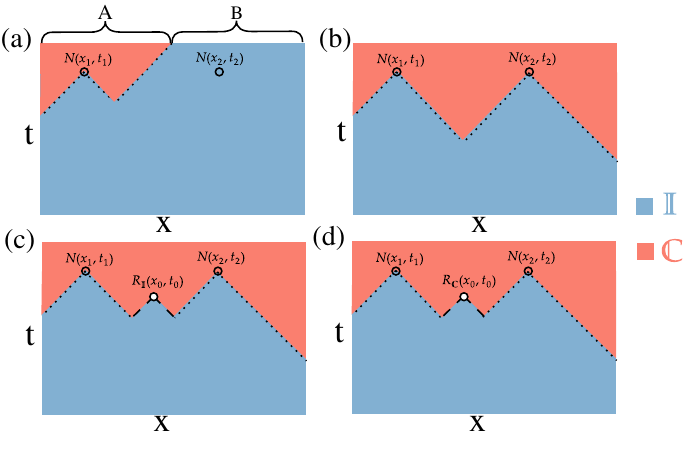}
\caption{\textit{Statistical model understanding of entanglement structure and information protection timescale}.
The schematic dominant spin configurations are illustrated here. The upper panel corresponds to the entanglement-generation setup, while the lower panel depicts the information-protection setup, where an additional Bell pair is introduced at $(x_{0}, t_{0})$. 
The horizontal ($x$) and vertical ($y$) axes represent spatial and temporal directions, respectively. 
Different colors denote distinct permutation-spin configurations in the effective classical model. 
The symbols $N$ and $R$ mark the locations of quantum-noise events and the inserted Bell pair, respectively; for clarity, additional noise events inside the $\mathbb{I}$ domain are omitted.
Panel~(a) shows the dominant configuration contributing to $F^{(n,k)}_{A}$. 
The presence of a midpoint defect on the top boundary modifies the local length scale of the adjacent domain-wall segment. 
Panel~(b) depicts the dominant configuration relevant for $F^{(n,k)}_{AB}$. 
Panels~(c) and (d) show the dominant configurations for $F^{(n,k)}_{AB}$ and $F^{(n,k)}_{AB \cup R}$, respectively, where the permutation spin at $(x_{0}, t_{0})$ is fixed to $\mathbb{I}$ or $\mathbb{C}$ depending on the boundary condition imposed by the Bell-pair encoding. In the presence of projective measurements, the domain wall exhibits fluctuations around its noise-determined trajectory. Reprinted with permission from Ref.~\cite{PhysRevLett.132.240402},
Copyright (2024) by the American Physical Society.}
\label{fig:DWunderstanding}
\end{figure*}

As illustrated in Fig.~\ref{fig:DWunderstanding}(b), spins remain in $\mathbb{C}$ until the backward-time propagation encounters a quantum noise $N(x_1,t_1)$ located at $(x_{1}, t_{1})$; all spins within the downward light-cone of this point flip from $\mathbb{C}$ to $\mathbb{I}$, while all others remain unchanged.  
Subsequent quantum noises inside this region no longer affect the configuration since the spins around these noises are already $\mathbb{I}$.  
A noise event outside the light-cone, such as $N(x_2,t_2)$, generates its own backward light-cone, converting its interior spins from $\mathbb{C}$ to $\mathbb{I}$.  
Thus, there is a domain wall separating $\mathbb{C}$ and $\mathbb{I}$ regions that is formed by the union of multiple light-cone boundaries, composed of many small segments.  
The typical distance of each small segment is governed by the mean spacing between noise events, yielding an effective length scale
\begin{equation}
    L_{\mathrm{eff}} \sim q^{-1}.
\end{equation}
As illustrated in Fig.~\ref{fig:DWunderstanding}(a), the dominant spin configuration for 
$\beta = A$ or $B$ can be interpreted in close analogy with that of $\beta = AB$. 
However, the distinct top boundary conditions imposed on $A$ and $B$ modify the local structure of the domain wall near the midpoint separating the two subsystems. 
In particular, this boundary constraint effectively reduces the characteristic length scale governing domain-wall fluctuations in this region. 
To leading order, the effective length scale is approximately halved, yielding an estimate of order $1/(2q)$ for the local domain-wall spacing.

Projective measurements act as a random Gaussian potential that perturbs the noise-determined domain-wall trajectory, inducing Kardar-Parisi-Zhang (KPZ) fluctuations~\cite{PhysRevLett.56.889,PhysRevLett.55.2923,PhysRevLett.55.2924}  around the original light-cone path.  
A domain wall of characteristic scale $q^{-1}$ therefore acquires a free-energy correction proportional to $q^{-1/3}$, in addition to its linear contribution.  
For the mutual information $I_{A:B}=S_{A} + S_{B}-S_{AB}$, bulk contributions cancel, leaving only the boundary free-energy cost near the midpoint between $A$ and $B$.  
Because this boundary contribution inherits the KPZ fluctuation behavior, we obtain the universal scaling as the leading linear terms also cancel:
\begin{equation}
    I_{A:B}(q) \sim q^{-1/3},
\end{equation}
which quantitatively characterizes the noise-induced area-law phase and demonstrates how quantum noise controls the steady-state entanglement structure.

\subsection{Temporally correlated bulk noise}
\label{subsec:Temporally correlated noise}
In the above discussion, we have established the $q^{-1/3}$ entanglement scaling in noisy monitored quantum circuits where noise events occur independently at random space--time locations. 
A natural question is whether temporal correlations between noise events can qualitatively alter this behavior. In the limit of strong temporal correlation, all noise events within a given trajectory share the same spatial locations across different discrete time steps, i.e., the spatial pattern of noises at every time slice is identical to that in the first time step shown in Fig.~\ref{fig:NoisyQCSetup} (b).

The theoretical framework introduced in Sec.~\ref{subsec:KPZ scaling from fluctuating domain walls}  can be naturally extended to the case of temporally correlated quantum noise. 
In this setting, correlated noise events act as emergent, extended boundaries in the statistical model, 
which in turn modify the dominant spin configuration and directly impose an effective length scale 
$L_{\mathrm{eff}} \sim q^{-1}$ and thus the $q^{-1/3}$ entanglement scaling of the steady state remains.

\subsection{Boundary noise}
\label{subsec:Boundary noise}
\begin{figure*}[t]
\centering
\includegraphics[width=0.7\textwidth, keepaspectratio]{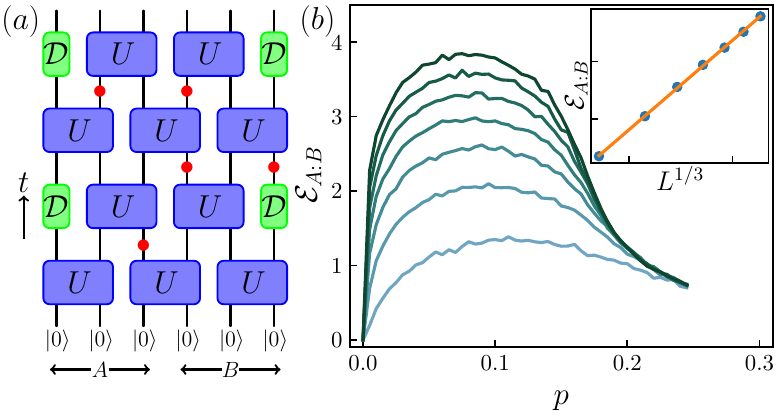}
\caption{\textit{Measurement-induced KPZ entanglement scaling}. (a) Noisy monitored quantum circuits with quantum noise located at the boundary. (b) The entanglement for the steady state scales as $L^{1/3}$. Reprinted with permission from Ref.~\cite{PhysRevLett.129.080501},
Copyright (2022) by the American Physical Society.}
\label{fig:BoundaryNoiseKPZ}
\end{figure*}

In addition to noisy monitored quantum circuits where quantum noises occur in the bulk, another important variant is the noisy monitored quantum circuit with quantum noises applied solely at the spatial boundaries~\cite{PhysRevLett.129.080501}. 
In each discrete time step, quantum noise acts on the boundary sites, while projective measurements are still applied in the bulk, as illustrated in Fig.~\ref{fig:BoundaryNoiseKPZ}. 
This setup exhibits two notable features: (i) the quantum noises are strongly temporally correlated, and (ii) the effective noise occurrence probability scales as $O(1/L)$. 
This motivates, and is closely connected to, our later discussion of bulk noise models with 
system-size-dependent occurrence probability (see Sec.~\ref{sec:Noise-induced phase transitions}). 
In the boundary-noise case, $q \sim 1/L$, implying an effective length scale $L_{\text{eff}} \sim L$. Consequently, the entanglement exhibits KPZ-type scaling $ L^{1/3}$, as shown in Fig.~\ref{fig:BoundaryNoiseKPZ}.

Throughout this discussion, the projective measurement probability is chosen such that 
$0 < p_{m} < p_{m}^{c}$ within the volume law phase, where $p_{m}^{c}$ is the critical measurement probability for MIPT 
in the absence of quantum noise. The main results on entanglement scaling of the steady state in noisy monitored quantum circuits are summarized in Table.~\ref{tab:EE}.

\begin{table*}
\begin{threeparttable}
\caption{Entanglement scaling in noisy monitored quantum circuits ($0<p_m<p_m^c$)}
\label{tab:EE}
\begin{tabular}{@{} l @{\hspace{1.2cm}} l @{}}
\toprule
 & Entanglement scaling of the steady state \\
\midrule
Temporally uncorrelated bulk noise & Area law, $q^{-1/3}$ \\
Temporally correlated bulk noise & Area law, $q^{-1/3}$ \\
Boundary noise & Power law, $L^{1/3}$ \\
\bottomrule
\end{tabular}
\end{threeparttable}
\end{table*}

\section{Information Protection Timescale of Steady States}
\label{sec:Information Protection in Noisy Monitored Circuits}
Although the entanglement scaling remains unchanged between temporally uncorrelated and temporally correlated quantum noises, the corresponding information-protection timescales differ markedly.
Specifically, we discuss the information-protection capability of the steady state in the presence of quantum noise. 
To probe this property, one qudit of quantum information is encoded into the steady state by creating a Bell pair between a reference qudit and a system qudit, as illustrated in Fig.~\ref{fig:NoisyQCSetup}.

\begin{figure*}[t]
\centering
\includegraphics[width=0.75\textwidth, keepaspectratio]{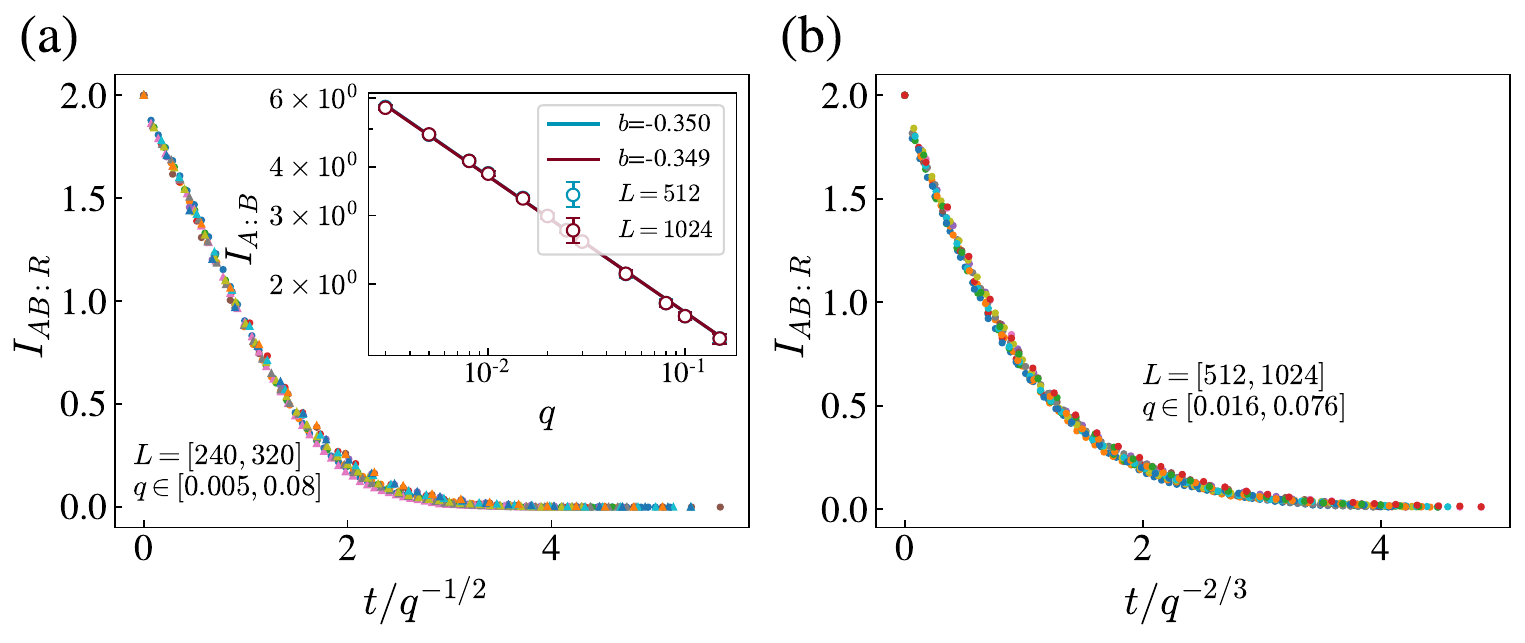}
\caption{\textit{Information protection dynamics}. $q$ represents the probability of reset channels and $p_{m}=0.2<p_{m}^{c}$ is the probability of projective measurement. (a) shows the mutual information $I_{AB : R}$ vs rescaled time $t/q^{-1/2}$ for temporally uncorrelated quantum noises. The inset shows the fitting of mutual information $I_{A:B}$ with the function $I_{A:B}(q)=aq^{b}$. $b$ is very close to the theoretical prediction $-1/3$. (b) shows the mutual information $I_{AB : R}$ vs rescaled time $t/q^{-2/3}$ for temporally correlated quantum noises. Reprinted with permission from Ref.~\cite{PhysRevLett.132.240402},
Copyright (2024) by the American Physical Society.}
\label{fig:Informationprotection}
\end{figure*}

\subsection{$q^{-2/3}$ scaling for temporally correlated noises and its relation with KPZ theory}
\label{subsec:temporally correlated noises and its relation with KPZ theory}

\begin{figure*}[t]
\centering
\includegraphics[width=0.75\textwidth, keepaspectratio]{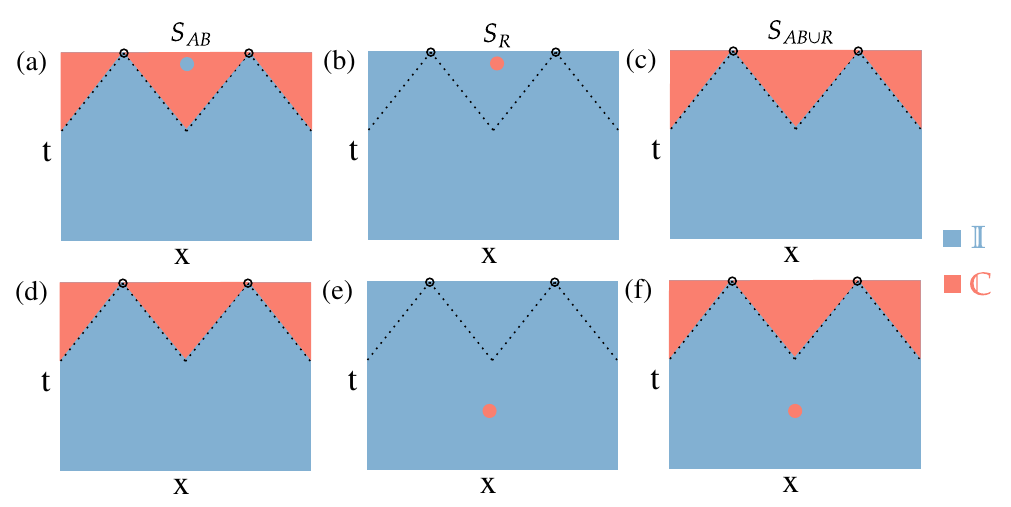}
\caption{\textit{Statistical model understanding of information protection timescale for temporally correlated noise}. The top and bottom panels correspond to the early- and late-time regimes, respectively. 
For $F^{(n,k)}_{AB}$, the permutation spin associated with the reference qubit and the system qubit forming a Bell pair with it is fixed to $\mathbb{I}$, resulting in a localized ``bubble'' (blue circle) in panel (a). 
As the system evolves in time, this bubble effectively moves downward. 
Once it crosses the domain wall separating the $\mathbb{C}$ and $\mathbb{I}$ domains, the dominant spin configuration changes from (a) to (d). The similar analysis can also be applied to the dominant spin configuration for $F^{(n,k)}_{AB \cup R}$ as shown in (c) and (f). Consequently, the vertical length scale of the domain wall therefore sets the information-protection timescale.
}
\label{fig:StatTemporalCorrelated}
\end{figure*}

The information-protection timescale for temporally correlated quantum noises can be naturally understood within the effective statistical-model description and is governed by KPZ-type domain-wall fluctuations. 
The dominant spin configurations contributing to $F^{(n,k)}_{AB}$, $F^{(n,k)}_{R}$, and $F^{(n,k)}_{AB \cup R}$ for $p_{m}=0$ are illustrated in Fig.~\ref{fig:StatTemporalCorrelated}, where the top and bottom panels correspond to early- and late-time regimes, respectively. 
Throughout Fig.~\ref{fig:StatTemporalCorrelated}, we set $p_{m}=0$; the inclusion of projective measurements ($0<p_{m}<p_{m}^{c}$) merely induces fluctuations of the domain wall and modifies its vertical length scale, without altering the underlying analytical picture.

As shown in Fig.~\ref{fig:StatTemporalCorrelated}(b,e), the dominant spin configuration governing $F^{(n,k)}_{R}$ consists of all spins fixed to $\mathbb{I}$ (blue region), except for the reference spin $R$, which is constrained to be $\mathbb{C}$ (red circle). 
Consequently, $S_{R}$ remains constant and receives contributions only from the localized ``bubble'' created by $R_{\mathbb{C}}$.

In contrast, the dominant spin configurations contributing to $F^{(n,k)}_{AB}$ and $F^{(n,k)}_{AB \cup R}$ are modified once the reference spin $R$ crosses the domain wall separating the $\mathbb{C}$ (red region) and $\mathbb{I}$ (blue region) domains. 
At this point, the mutual information $I_{AB:R}$ abruptly changes from $2$ to $0$, signaling a transition from perfect information protection to complete information loss.
Accordingly, the information-protection timescale is determined by the average height of the domain wall---scaling as $q^{-1}$ in the absence of projective measurements and as $q^{-2/3}$ when monitored measurements induce KPZ-type fluctuations with wandering exponent $\chi=2/3$. The numerical simulation has verified this analytical prediction, as shown in Fig.~\ref{fig:Informationprotection}(b).

As discussed in Sec.~\ref{subsec:Boundary noise}, boundary noise applied at fixed spatial boundaries can be regarded as a special case of temporally correlated bulk noise. In this situation, the effective length scale scales as $L_{\text{eff}} \sim L$, leading to a steady-state information-protection timescale of $O(L^{2/3})$. Another special case arises when quantum noise acts only at one single spatial boundary instead of both; see Sec.~\ref{subsec:Noise-induced coding transition} for a detailed discussion. 

\begin{figure*}[t]
\centering
\includegraphics[width=0.75\textwidth, keepaspectratio]{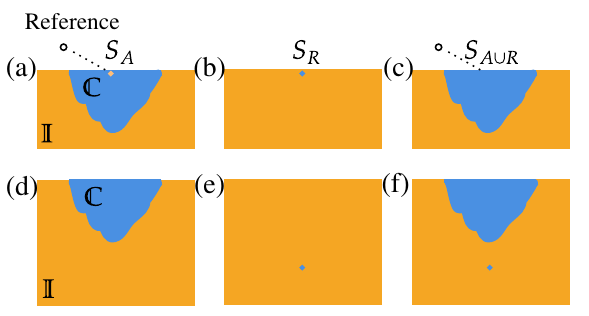}
\caption{\textit{Statistical model understanding of the subsystem information-protection timescale of MIPTs}. 
Similar to the case with temporally correlated noise, the subsystem information-protection timescale is set by the vertical domain-wall length scale. 
Consequently, it scales as $O(L_{A})$ for $p_{m}=0$ and as $O(L_{A}^{2/3})$ for $0<p_{m}<p_{m}^{c}$. Reprinted with permission from Ref.~\cite{PhysRevLett.132.240402},
Copyright (2024) by the American Physical Society.}
\label{fig:MIPT1}
\end{figure*}

Moreover, we note that this analytical understanding can be straightforwardly generalized to subsystem information protection in the steady state of conventional MIPTs. In this setting, instead of considering the mutual information $I_{AB:R}$ between the full system $AB$ and a reference qudit $R$, we focus on the mutual information $I_{A:R}$, where $A$ is a subsystem containing the system qudit that forms a Bell pair with the reference qudit during the encoding process. When the size of subsystem $A$ is smaller than half of the total system, i.e., $L_{A} < L/2$, the remaining degrees of freedom can be regarded as an effective bath or environment. In this case, the statistical-model description is identical to that for temporally correlated noise. As illustrated in Fig.~\ref{fig:MIPT1}, once the reference spin $R$ crosses the domain wall, the dominant spin configurations governing $F^{(n,k)}_{A}$ and $F^{(n,k)}_{A \cup R}$ are altered, causing the mutual information to drop from $2$ to $0$. Consequently, the information-protection timescale is controlled by the vertical length scale of the domain wall and scales as $O(L_{A})$ when $p_{m}=0$, and as $O(L_{A}^{2/3})$ when $0 < p_{m} < p_{m}^{c}$ in the volume law phase.

\subsection{$q^{-1/2}$ scaling for temporally uncorrelated noises and its relation with Hayden-Preskill protocol}
\label{subsec:temporally uncorrelated noises and its relation with Hayden-Preskill protocol}

However, for temporally uncorrelated noise, the information-protection timescale is no longer governed by the vertical domain-wall length scale. 
In the temporally correlated case, the reference qudit effectively moves downward in time without encountering additional noise events, and the dominant spin configuration changes only when the reference qudit crosses the domain wall. 
By contrast, for temporally uncorrelated noise, the random space--time distribution of noise events allows the effectively moving reference qudit to encounter noise before reaching the domain wall, thereby altering the dominant spin configuration even before the original domain wall is crossed. See Fig.~\ref{fig:DWunderstanding}(c,d) for an alternative dominant spin configuration. 
At early time after encoding, where the reference space--time position plays a role similarly to a noise. 
For a longer time scale, when there are several noise events in the red region, they can form the new domain wall and leave the reference space--time position in the $\mathbb{I}$ region.

\begin{figure*}[t]
\centering
\includegraphics[width=0.65\textwidth, keepaspectratio]{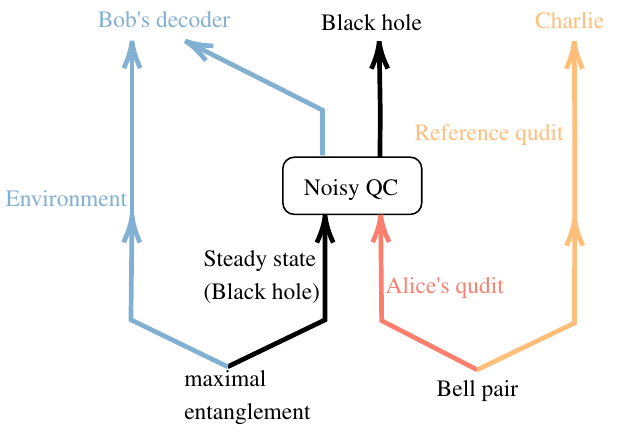}
\caption{\textit{Connection between information protection in noisy quantum circuits and Hayden-Preskill protocol.} The steady state can be regarded as a black hole and Alice throws one-qudit information into the black hole to destroy it. The timescale of information protection corresponds to the time required for Bob to decode the information from collecting the qudits released by Hawking radiation. Reprinted with permission from Ref.~\cite{PhysRevLett.132.240402},
Copyright (2024) by the American Physical Society.}
\label{fig:HaydenPreskill}
\end{figure*}

This behavior can be understood more intuitively through the Hayden--Preskill protocol~\cite{PatrickHayden_2007}, 
schematically depicted in Fig.~\ref{fig:HaydenPreskill}. 
Here the steady state plays the role of an old black hole maximally entangled with its environment (Bob). 
Alice encodes a one-qudit quantum message, maximally entangled with Charlie’s reference qudit, by injecting it into the black hole. 
Quantum noise channels function analogously to Hawking radiation emitted during the evaporation process. 
According to the Hayden--Preskill protocol, Bob needs only slightly more than one qudit of Hawking radiation to reconstruct Alice’s message. 
Consequently, the information-protection timescale is set by the time required for a noise event, occurring with probability $q$, to fall within the backward light-cone of area $O(t^{2})$, leading to the scaling
\begin{equation}
    t \sim q^{-1/2}.
\end{equation}
The presence of projective measurements does not modify this qualitative understanding. 
Thus, the information-protection timescales for temporally uncorrelated quantum noise scale as $q^{-1/2}$ for both $p_{m}=0$ and $0<p_{m}<p_{m}^{c}$.

Consequently, temporal correlations in the quantum noise lead to qualitatively distinct information-protection timescales in the steady state of noisy monitored circuits. In Sec.~\ref{subsec:Noise-induced coding transition}, we turn to the complementary setting of initial-state encoding. 
The main results on the information protection timescales in noisy monitored quantum circuits are summarized in Table.~\ref{tab:Information}.

\begin{table*}
\begin{threeparttable}
\caption{Information protection timescales with different settings ($0<p_{m}<p_{m}^{c}$).}
\label{tab:Information}
\begin{tabular}{@{} l @{\hspace{1.2cm}} l @{\hspace{1.2cm}} l @{}}
\toprule
 & Initial state (product state) & \textcolor{magenta}{Steady state} \\
\midrule
Subsystem ($L_{A}<L/2$) without noise
& $O(L_{A})$
& \textcolor{magenta}{$O(L_{A}^{2/3})$}
\\[0.6ex]
Whole system with temporally correlated noise
& $O(q^{-1})$
& \textcolor{magenta}{$O(q^{-2/3})$}
\\[0.6ex]
Whole system with temporally uncorrelated noise
& $O(q^{-1})$
& \textcolor{magenta}{$O(q^{-1/2})$}
\\
\bottomrule
\end{tabular}
\end{threeparttable}
\end{table*}

These intriguing dynamical phenomena including entanglement structure and information protection dynamics, uncovered in noisy monitored quantum circuits further motivate the study of novel quantum dynamics in broader classes of open quantum systems, such as monitored Hamiltonian systems with decoherence~\cite{PhysRevResearch.2.013022,PhysRevLett.127.140601,PhysRevB.103.224210,NonlocalMIPT_Quantum,PhysRevB.107.245132}. From an experimental perspective, these models provide compelling and accessible simulation tasks, potentially enabling the discovery of new nonequilibrium phases of matter.

However, in monitored systems, the intrinsic randomness of projective measurement outcomes gives rise to a severe post-selection problem~\cite{noelMeasurementinducedQuantumPhases2022b, kohMeasurementinducedEntanglementPhase2023, hokeMeasurementinducedEntanglementTeleportation2023}, which poses a major challenge for experimental investigations of emergent phenomena. In particular, the probability of obtaining identical measurement trajectories required for measuring entanglement decreases exponentially with the number of measurements, leading to an exponentially large experimental cost. 
To address this challenge, several strategies have been proposed, including the use of cross-correlation~\cite{PRXQuantum.5.030311}, driven dynamics based on the Kibble--Zurek mechanism~\cite{wang2024drivencriticaldynamicsmeasurementinduced}, and machine-learning-based approaches~\cite{hou2025machinelearningeffectsquantum, kim2025learningmeasurementinducedphasetransitions}. 
These developments substantially mitigate the post-selection problem, making the experimental exploration of novel phenomena in monitored systems increasingly feasible.

\section{Noise-induced phase transitions}
\label{sec:Noise-induced phase transitions}
In the preceding discussion, the occurrence probability of quantum noise was assumed to be independent of the system size. 
We now consider a more general setting in which
\begin{equation}
    q = \frac{p}{L^{\alpha}},
\end{equation}
where $\alpha = 0$ recovers the system-size-independent case analyzed above.  
In this broader framework, we discuss the resulting noise-induced phase transitions. We note that the setting involving boundary-localized noise~\cite{Coding_Vijay} corresponds to a special case of this general framework with $\alpha = 1$.

\subsection{Noise-induced entanglement phase transition}
\label{subsec:Noise-induced entanglement phase transition}
Although no entanglement phase transition occurs in the case $\alpha = 0$, 
this more general framework allows noise-induced phase transitions to emerge as $p$ changes with $\alpha=1$. 
We set $T=O(L)$, the same setting as that in normal MIPTs.

\begin{figure*}[t]
\centering
\includegraphics[width=0.7\textwidth, keepaspectratio]{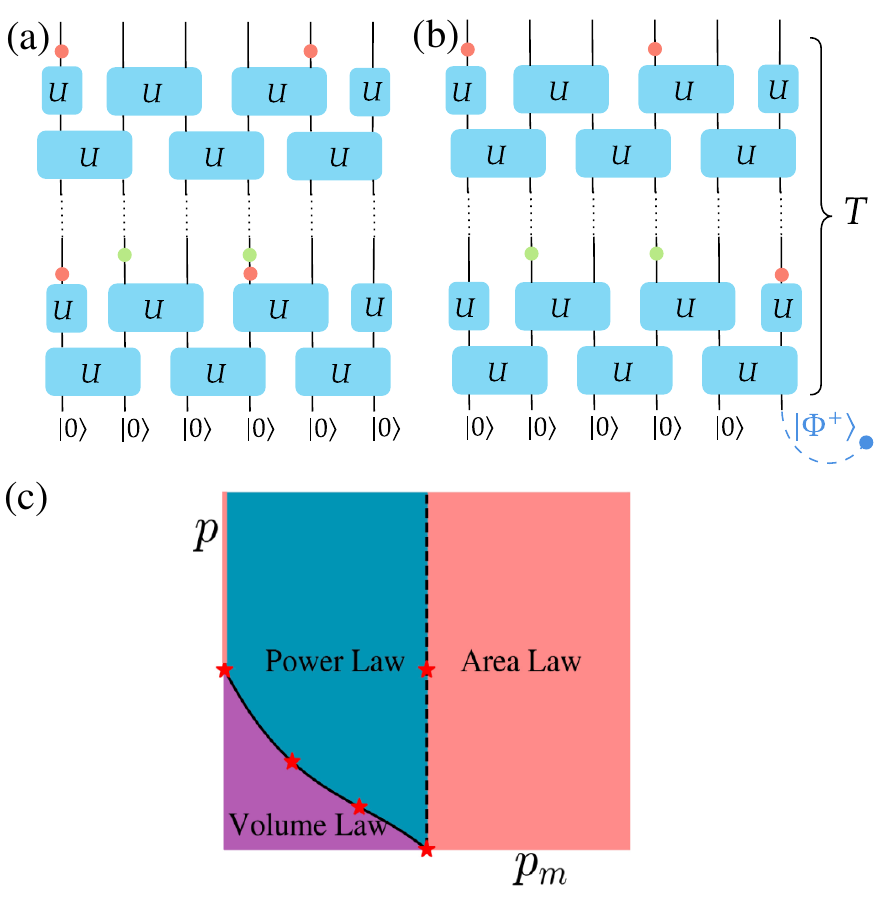}
\caption{\textit{Setup and phase diagram for noise-induced entanglement and coding phase transitions.} Circuit setups with $6$ qudits for (a) entanglement phase transition and (b) coding transition. 
The red and green circles represent the quantum channels and projective measurements, respectively. 
In (b), a qudit is maximally entangled with a reference qudit by creating a Bell pair $\vert \Phi^{+} \rangle$ to encode one qudit information. 
(c) Phase diagram of the entanglement phase transition with $T=4L$. 
Red stars represent the critical points identified from numerical results. 
The black solid (dashed) curve denotes the noise (measurement)-induced phase transition. Reprinted with permission from Ref.~\cite{PhysRevB.110.064323},
Copyright (2024) by the American Physical Society.}
\label{fig:Coding}
\end{figure*}

Again, we first provide theoretical understanding from the statistical model.
For the classical spin model associated with $S_{AB}$, the dominant configuration for $\alpha > 1$ is the fully aligned state in which all spins are fixed to $\mathbb{C}$, identical to the noiseless case. 
The free energy of this configuration scales as $qLT$, the average number of quantum-noise events, reflecting the magnetic-field--like pinning effect of the noise.  
In contrast, when $\alpha < 1$, the fully-$\mathbb{C}$ configuration is no longer energetically favored.  
Instead, the configuration containing a domain wall, as discussed in Sec.~\ref{subsec:KPZ scaling from fluctuating domain walls}, has lower free energy, which scales as $s_{0}L$, where $s_{0}$ is a constant. 
Moreover, as discussed in Sec.~\ref{subsec:KPZ scaling from fluctuating domain walls}, the domain wall inherits an effective length scale $L_{\mathrm{eff}} \sim q^{-1}$.

For the classical spin model corresponding to $S_{A}$ or $S_{B}$, the domain-wall configuration is always energetically preferred due to the fixed top boundary condition.  
Consequently, when $\alpha = 1$, the system exhibits a competition between the two candidate configurations for $S_{AB}$.  
This leads to a volume-law entanglement phase when $p < s_{0}L/T$, with mutual information scaling as $(s_{0} - pT/L)L$, and an area-law phase when $p > s_{0}L/T$.  
This crossover corresponds to the vertical transition line at $p_{m} = 0$ in Fig.~\ref{fig:Coding}(c).

Projective measurements do not modify the basic mechanism of the noise-induced transition—namely, 
the competition between the two spin configurations.  
However, they act as an attractive random Gaussian potential in the statistical model, 
causing the domain wall to fluctuate in such a way that it passes through more measurement events, thereby lowering its free energy.  
Following Refs.~\cite{PhysRevB.107.L201113, PhysRevLett.132.240402}, 
the resulting domain-wall free energy acquires KPZ-type scaling, with a subleading correction proportional to 
$L_{\mathrm{eff}}^{1/3} \sim L^{1/3}$ ($\alpha=1$).  
Therefore, for $p > p_{c}$, the entanglement exhibits a power-law scaling $L^{1/3}$ in the presence of projective measurements, 
as indicated by the blue region in Fig.~\ref{fig:Coding}(c).

Importantly, this entanglement transition is a first-order transition arising from the competition between two distinct 
spin configurations.  
For $\alpha > 1$ ($\alpha < 1$), the fully-$\mathbb{C}$ configuration (the domain-wall configuration) always dominates, 
yielding a single volume-law (or power-law/area-law) phase.  
Thus, the noise-induced entanglement transition occurs only at the marginal point $\alpha = 1$.

\subsection{Noise-induced coding transition}
\label{subsec:Noise-induced coding transition}
In Sec.~\ref{sec:Information Protection in Noisy Monitored Circuits}, we have discussed the information protection timescale of the steady state of the noisy monitored quantum circuits. In this section, we further discuss the information protection capacity of the initial state as well as the noise-induced coding transition.

We now extend the statistical-model framework to analyze the noise-induced coding transition. 
In addition to the top boundary conditions discussed previously, the spin at the bottom boundary, corresponding to the location where the Bell pair is introduced, is fixed by the boundary condition of the reference qudit: 
$\mathbb{I}$, $\mathbb{C}$, and $\mathbb{C}$ for $S_{AB}$, $S_{R}$, and $S_{AB \cup R}$, respectively  

For $S_{R}$, the dominant configuration features a bulk pinned to $\mathbb{I}$ independently of the noise probability.  
Thus, $S_{R}$ is constant, receiving only a unit contribution from the localized defect created by the Bell pair.  

\begin{figure*}[t]
\centering
\includegraphics[width=0.75\textwidth, keepaspectratio]{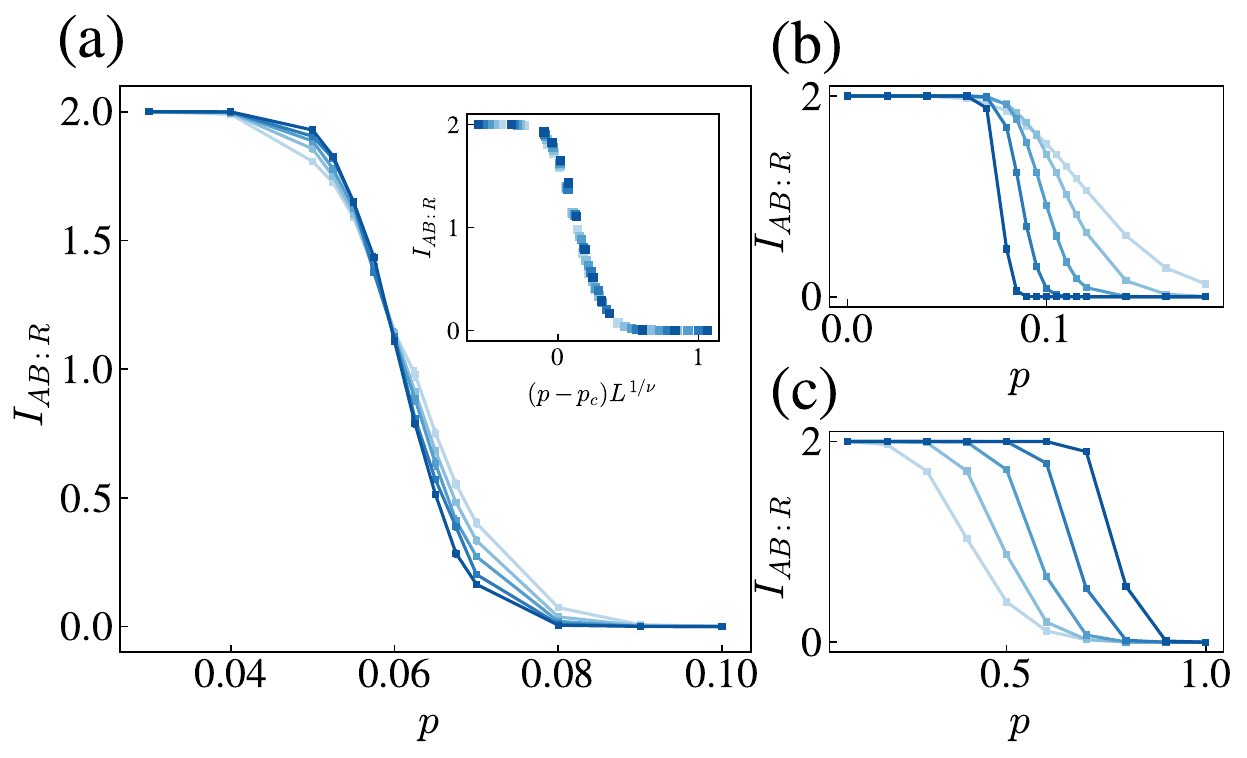}
\caption{\textit{Numerical results of noise-induced coding transition}. (a) Quantum noise with $\alpha = 1$, $p_{m}=0.2$, and $T = 4L$. Below $p_{c}$, the encoded information is perfectly protected.   
(b),(c) Mutual information $I_{AB:R}$ for $\alpha = 0.8$ and $\alpha = 1.2$, respectively.  
The noise-induced phase transition disappears in the thermodynamic limit when $\alpha \neq 1$. Reprinted with permission from Ref.~\cite{PhysRevB.110.064323},
Copyright (2024) by the American Physical Society.}
\label{fig:BulkCoding}
\end{figure*}

In contrast, for $S_{AB}$ and $S_{AB \cup R}$, the competition persists between two distinct spin configurations: 
(i) the fully aligned configuration with all spins fixed to $\mathbb{C}$, and  
(ii) the configuration containing a topmost domain wall, similar to that discussed in Sec.~\ref{subsec:Noise-induced entanglement phase transition}.  
When the former configuration dominates, the mutual information is $I_{AB:R}=2$, indicating perfect protection of the encoded information.  
When the latter dominates, $I_{AB:R}=0$, signaling complete loss of the encoded information.  
Thus, the encoded information remains perfectly protected below a critical value of the noise-strength prefactor $p$, and a noise-induced coding transition occurs as $p$ increases.  
Within the unified statistical-model description, this coding transition is shown to share the same critical point and critical exponent as the noise-induced entanglement phase transition, which has been verified numerically as shown in Fig.~\ref{fig:BulkCoding}. Moreover, in this initial-state encoding setup, temporal correlations between quantum noise events do not change the information-protection timescale, in contrast to the behavior in the steady-state encoding setup.

\begin{figure*}[t]
\centering
\begin{subfigure}{0.45\textwidth}
    \centering
    \includegraphics[width=\textwidth]{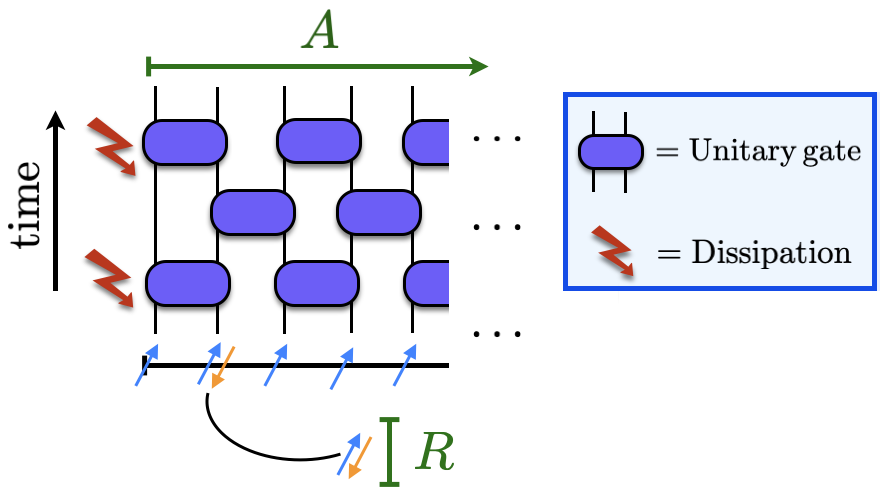}
\end{subfigure}
\hspace{0.02\textwidth}
\begin{subfigure}{0.45\textwidth}
    \centering
    \includegraphics[width=\textwidth]{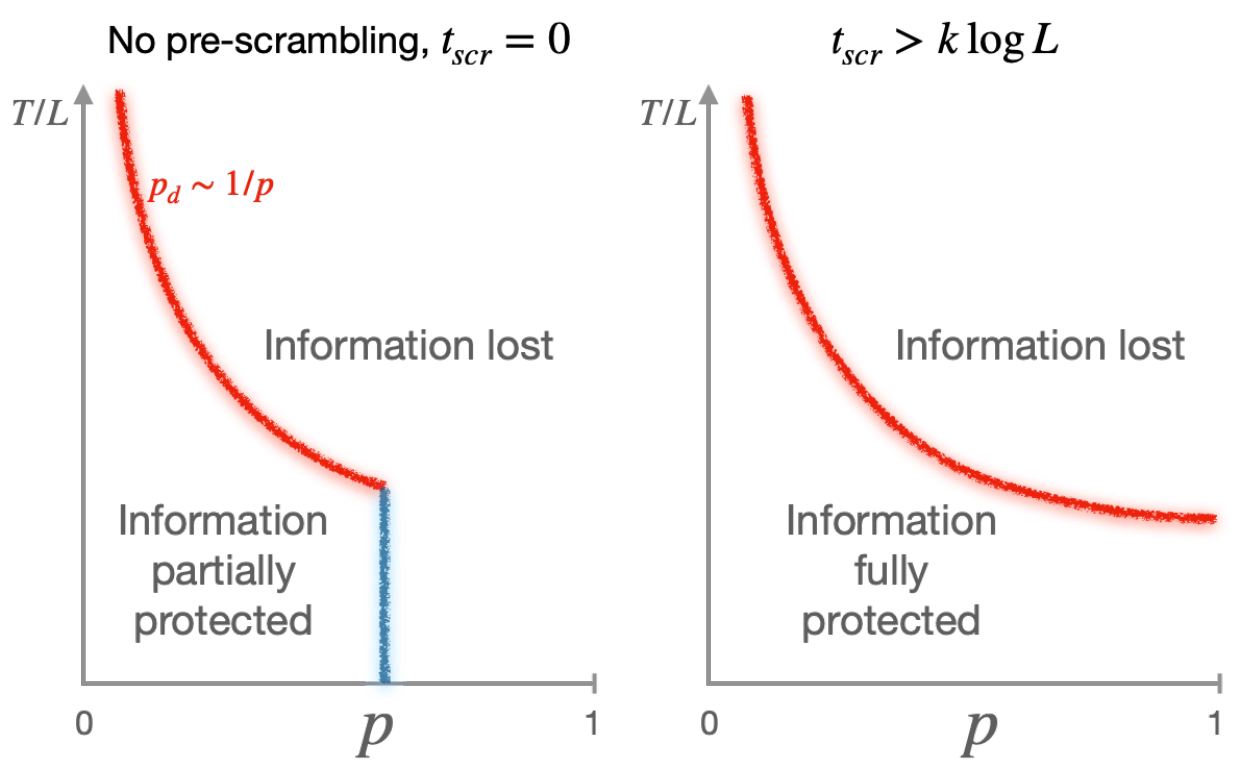}
\end{subfigure}
\caption{\textit{Setup and phase diagram of boundary noise-induced coding transition}. 
Left: schematic setup in which quantum noise is applied only at the left boundary. 
Right: phase diagram. 
Below the critical noise probability $p_{c}$, the encoded information is only \emph{partially} protected, in contrast to the perfect protection in the case with bulk noise. 
Notably, when $T/L > 1$, the coding transition is first order (red line), whereas it becomes continuous when $T/L < 1$. 
With an additional pre-scrambling evolution, the encoded information can be perfectly protected. Reproduced from Ref.~\cite{Coding_Vijay}. \href{https://creativecommons.org/licenses/by/4.0/}{CC BY 4.0}.}
\label{fig:BoundaryCoding}
\end{figure*}

In Ref.~\cite{Coding_Vijay}, the noise-induced coding transition with quantum noise applied only at the left spatial boundary was investigated. The setup and corresponding phase diagram are shown in Fig.~\ref{fig:BoundaryCoding}. This boundary-noise-induced coding transition differs qualitatively from the bulk-noise case. On the one hand, when the noise probability is below the critical value $p_{c}$, the encoded information is \emph{perfectly} protected ($I_{AB:R}=2$) in the presence of bulk noise, whereas it is only \emph{partially} protected ($0 < I_{AB:R} < 2$) when the noise acts solely at the left boundary. On the other hand, for bulk noise the coding transition is always first order. 
In contrast, for boundary noise the transition is first order when $T/L > 1$, matching the behavior of the bulk case, but becomes continuous when $T/L < 1$. More detailed discussions can be found in Refs.~\cite{Coding_Vijay, PhysRevB.110.064323}. 
We note that the analytical understanding based on the effective statistical model in the large-$d$ (local Hilbert space) limit does not qualitatively match numerical results obtained at finite $d$. 
This discrepancy arises because, at finite $d$ (finite temperature for statistical model), there is an additional entropic contribution to the free energy for the boundary noise case. This entropic effect plays a crucial role in the boundary-noise-induced coding transition at finite $d$. 
See Ref.~\cite{PhysRevB.110.064323} for a more detailed discussion.

Moreover, this initial state encoding and information protection scheme can also be extended to detect subsystem information protection where the complementary system is regarded as the environment. In such settings, the subsystem information protection measure is called the subsystem information capacity, which can be used to probe and characterize different dynamical phases~\cite{Chen2025subsystem, Qing2025z}.

\subsection{Noise-induced complexity transition}
\label{subsec: noise-induced complexity phase transition}
\begin{figure*}[t]
\centering
\includegraphics[width=0.49\textwidth, keepaspectratio]{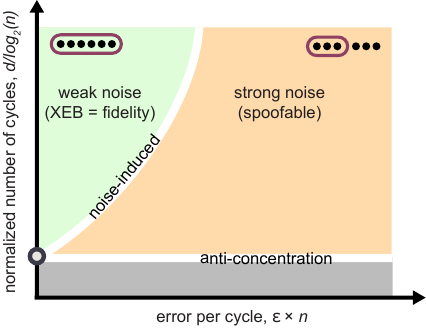}
\caption{\textit{Noise-induced complexity transition in random circuit sampling}. In the weak-noise regime, the system is not classically simulable, whereas in the strong-noise regime it becomes classically simulable. Reproduced from Ref.~\cite{morvanPhaseTransitionsRandom2024}. \href{https://creativecommons.org/licenses/by-nc-nd/4.0/}{CC BY-NC-ND 4.0}.}
\label{fig:RCS}
\end{figure*}

In addition to noise-induced entanglement and coding transitions, a noise-induced complexity phase transition in random circuit sampling has also been identified~\cite{NIPT2_arxiv, PhysRevA.109.042414} and experimentally demonstrated on a superconducting quantum processor~\cite{morvanPhaseTransitionsRandom2024}. 
The boundary between these phases can be resolved in finite-size systems using a fidelity-estimation technique known as cross-entropy benchmarking~\cite{boixoCharacterizingQuantumSupremacy2018a,doi:10.1126/science.aao4309,QI_nature19}. 
When the quantum noise is strong, the wavefunction effectively factorizes into multiple weakly correlated subsystems, and the circuit output can be efficiently spoofed by classical algorithms that simulate only a subset of the system. 
Conversely, when the noise is sufficiently weak, correlations extend across the entire circuit, restoring the intrinsic computational complexity required for quantum sampling. 
The phase diagram of this noise-induced complexity phase transition is shown in Fig.~\ref{fig:RCS}.
We note that, similar to the noise-induced entanglement and coding transitions discussed above, this complexity transition also disappears when $\alpha \neq 1$.

\subsection{Connections and Distinctions Between Noise-Induced Phase Transitions}
In this section, we have introduced three types of noise-induced phase transitions: the noise-induced entanglement transition, coding transition, and complexity transition in random circuit sampling. 
All of these transitions occur when $\alpha = 1$ in the noise probability and disappear for other choices of $\alpha$. 
From the perspective of the effective statistical model, the noise-induced entanglement and coding transitions share the same critical exponents and critical points that have been verified numerically. 
Moreover, their critical noise probabilities both depend on the ratio $L/T$ and vanish in the limit $L/T \to 0$ even when $\alpha = 1$, consistent with the fact that any encoded information is ultimately destroyed in the presence of quantum noise.

In contrast, the noise-induced complexity transition in random circuit sampling~\cite{morvanPhaseTransitionsRandom2024, NIPT2_arxiv, PhysRevA.109.042414}, which is characterized by the linear cross-entropy benchmark rather than entanglement-based observables, exhibits a qualitatively different statistical-model structure. Only two replicas are involved in the statistical model analysis, and the critical noise probability $p_{c}$ is independent of the ratio $L/T$, with analytical predictions yielding $p_{c} \approx 1$.

\section{Broader Applications of Noisy Quantum Circuits}
\label{sec:Broader Applications of Noisy Quantum Circuits}

Beyond their intrinsic theoretical interest, noisy quantum circuits have inspired a wide range of constructions and applications across quantum information science, quantum computation, and quantum many-body physics.
In this section, we discuss broader applications of noisy quantum circuits.

\subsection{Variational quantum algorithms}
Solving for the ground state of a given many-body Hamiltonian is a central task in quantum many-body physics. In the noisy intermediate-scale quantum (NISQ) era, the variational quantum eigensolver (VQE)~\cite{peruzzoVariationalEigenvalueSolver2014c,kandalaHardwareefficientVariationalQuantum2017b,cerezoVariationalQuantumAlgorithms2021b,Tilly_2022,PhysRevLett.128.120502,PhysRevB.107.024204,PhysRevLett.131.073602,PhysRevLett.132.150603,PhysRevResearch.5.L032040} is one of the most promising algorithms to get the target ground state, in which a parameterized quantum circuit is optimized by a classical optimizer to minimize a cost function, usually chosen to be the energy.

However, the barren plateau problem~\cite{mccleanBarrenPlateausQuantum2018b} has severely limited the applicability of VQE, since for generic parameterized circuits the gradient of the cost function with respect to the variational parameters vanishes exponentially with system size. The presence of barren plateau in noiseless parameterized circuits is linked to random parameter initialization. Moreover, in the NISQ era, noise must also be taken into account when investigating the trainability of VQE. In the presence of noise (decoherence), namely in noisy parameterized quantum circuits, noise-induced barren plateaus have been identified as a generic obstacle to trainability~\cite{wangNoiseinducedBarrenPlateaus2021,Schumann_2024}. In this case, the gradient vanishes with increasing problem size at every point on the cost-function landscape, rather than only probabilistically as in noiseless parameterized circuits. Therefore, developing new strategies to mitigate noise-induced barren plateaus is a central challenge for applying VQE on NISQ devices.

Quantum channels can be broadly classified into two types. Non-unital quantum channels do not preserve the identity operator, i.e., \(\mathcal{E}(I)\neq I\). For example, under the action of a quantum reset channel, one has \(\mathcal{E}(I)=\ket{0}\bra{0}\). In contrast, unital quantum channels satisfy \(\mathcal{E}(I)=I\). The quantum dephasing channel and the depolarizing channel are examples of unital channels. Interestingly, the presence of non-unital quantum channels can mitigate or even eliminate barren plateaus~\cite{mele2024noiseinducedshallowcircuitsabsence}. This can be understood from the fact that non-unital channels effectively reduce the circuit depth and it is known that shallow circuits of finite or logarithmic depth are free from barren plateaus~\cite{cerezoCostFunctionDependent2021b,Uvarov_2021}, whereas unital channels tend to drive the output state of a noisy parameterized quantum circuit toward a maximally mixed state, thereby causing the gradients to vanish.

\begin{figure*}[t]
\centering
\includegraphics[width=0.65\textwidth, keepaspectratio]{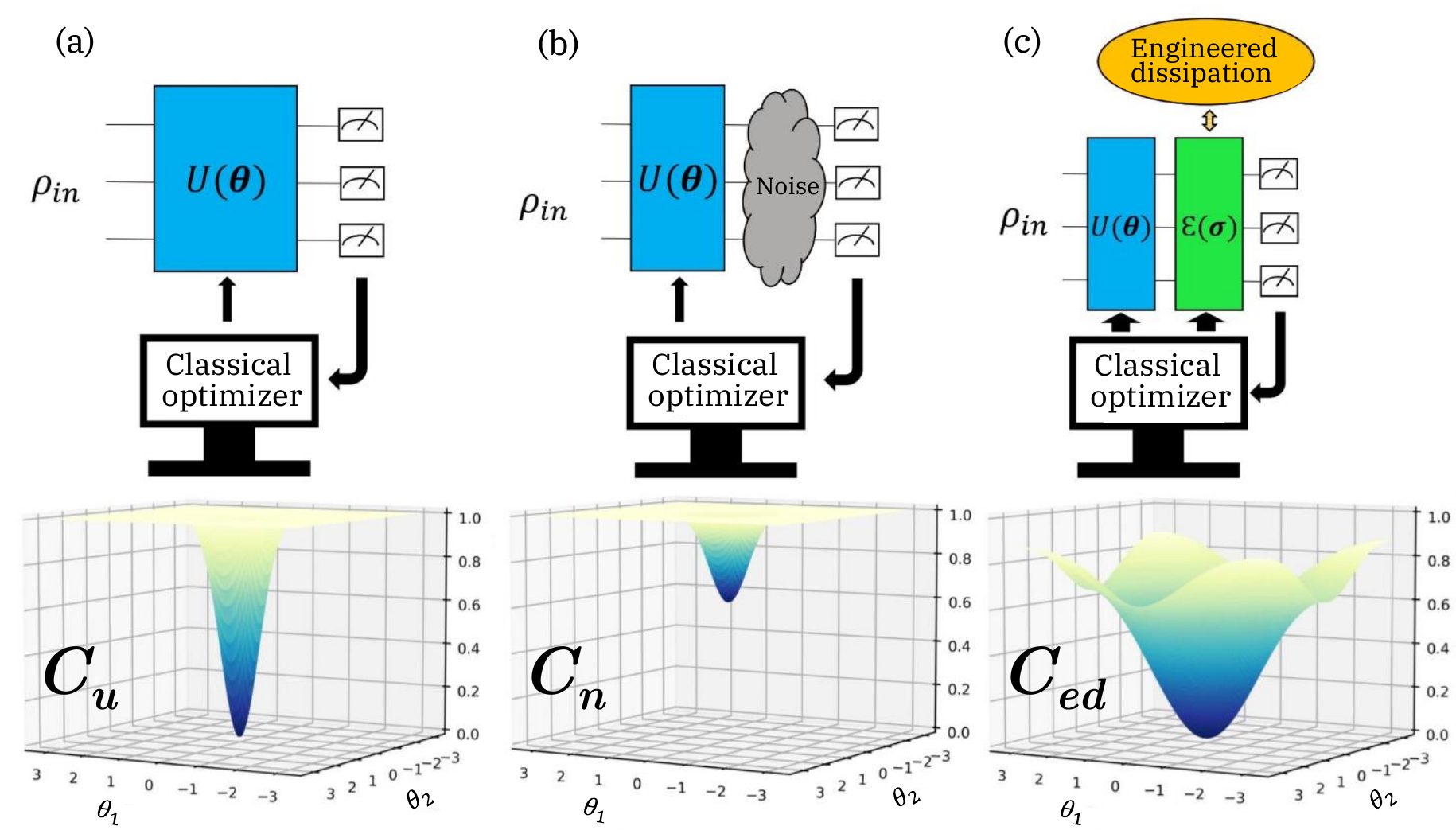}
\caption{\textit{Comparison of optimization landscapes in noisy quantum circuits}. (a) fully unitary ansatz, (b) ansatz with depolarizing noise channels, (c) engineered-dissipation ansatz. The engineered-dissipation ansatz has the maximum variance and thus mitigate barren plateaus. Reproduced from Ref.~\cite{sanniaEngineeredDissipationMitigate2024}. \href{https://creativecommons.org/licenses/by/4.0/}{CC BY 4.0}.
}
\label{fig:NoiseMitagteBP}
\end{figure*}

Moreover, properly engineered dissipative mechanisms can also mitigate barren plateaus~\cite{sanniaEngineeredDissipationMitigate2024}. In particular, introducing suitably designed Markovian non-unitary layers into parameterized quantum circuits can improve trainability, because a global optimization problem can be effectively mapped onto a more local one. See Fig.~\ref{fig:NoiseMitagteBP} for a comparison of the optimization landscapes in noiseless parameterized circuits, generic noisy parameterized circuits, and noisy parameterized circuits with engineered dissipation.

These findings highlight that noise, rather than being purely detrimental, can in some cases serve as a resource for improving optimization landscapes and enhancing the performance of VQE on NISQ devices. Therefore, guiding the design of noise-benefited variational ansatzes is an important direction in the NISQ era.

\subsection{Classical simulation of noisy quantum circuits}

Noisy quantum circuits have also played a central role in sharpening our understanding of the boundary between classical simulability and genuine quantum advantage. While the computational cost of simulating noiseless RQCs grows exponentially, the introduction of noise can drastically alter their complexity. As discussed in Sec.~\ref{subsec: noise-induced complexity phase transition}, there is a noise-induced complexity phase transition that separates a regime that remains classically intractable from one that becomes 
efficiently simulable~\cite{NIPT2_arxiv, PhysRevA.109.042414,morvanPhaseTransitionsRandom2024}. 

\begin{figure*}[t]
\centering
\includegraphics[width=0.5\textwidth, keepaspectratio]{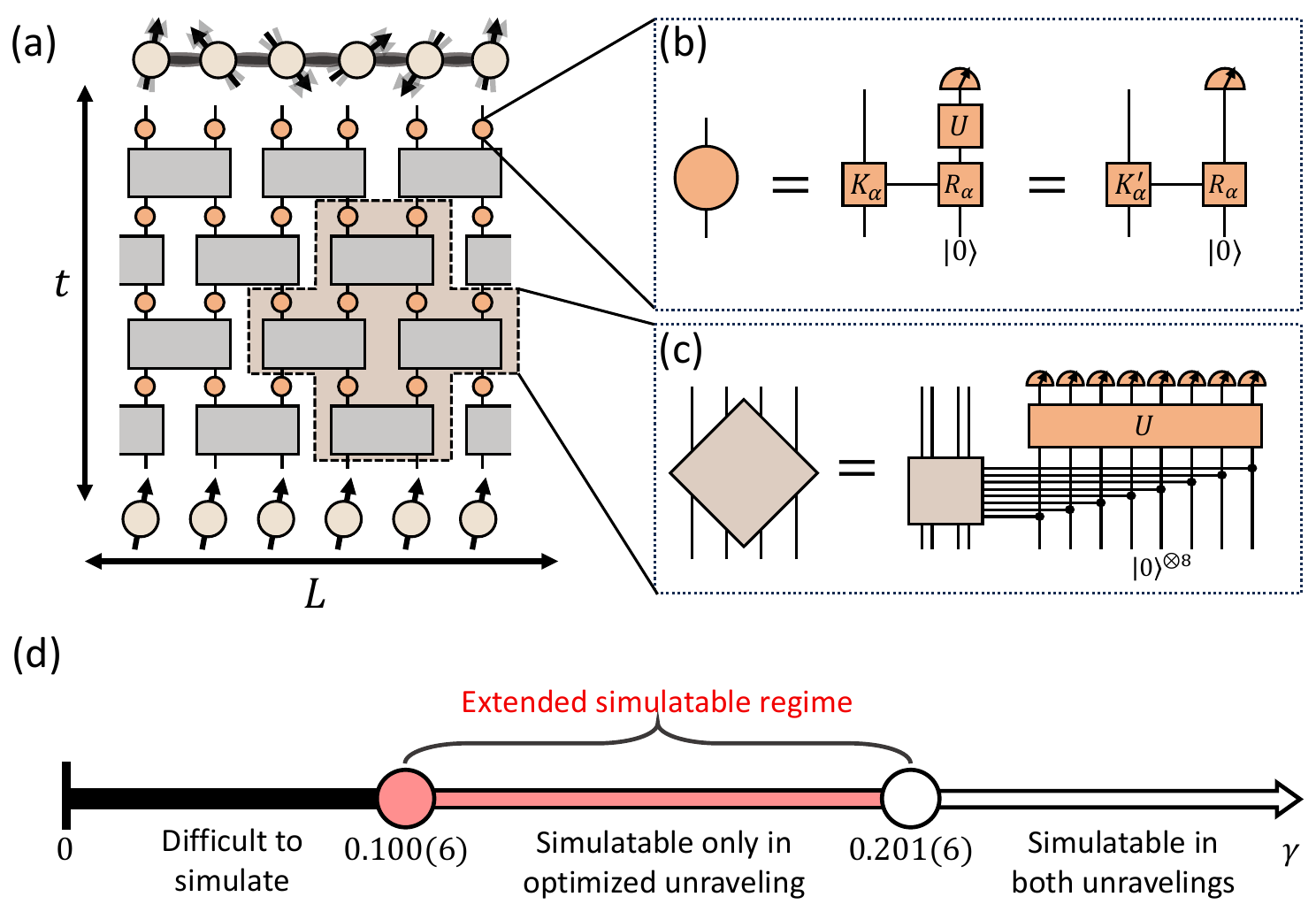}
\caption{\textit{Illustration of trajectory unraveling for noisy quantum circuits}. There is an extended simulable regime obtained using 
trajectory unraveling.
Reproduced from Ref.~\cite{PhysRevLett.133.230403}. Copyright (2024) by the American Physical Society.
}
\label{fig:unravelingFig1}
\end{figure*}

This line of inquiry has additionally motivated the development of classical simulation techniques specifically tailored for noisy circuits. 
When the noise rate exceeds a certain threshold, the real-time dynamics of noisy quantum circuits can be efficiently simulated using matrix product operators~\cite{Noh2020efficientclassical} or tensor-network formalisms~\cite{gao2018efficientclassicalsimulationnoisy}. More recently, unraveling-based trajectory methods~\cite{PRXQuantum.4.040326,PhysRevLett.133.230403,cichy2025classicalsimulationnoisyquantum} and Pauli-propagation or path-integral-inspired approaches~\cite{10.1145/3564246.3585234,PhysRevLett.133.120603,angrisani2025simulatingquantumcircuitsarbitrary,lee2025classicalsimulationnoisyrandom,j1gg-s6zb,GonzalezGarcia2025paulipath,li2025dualrolelowweightpauli,fontana2023classicalsimulationsnoisyvariational} have also been proposed for simulating noisy quantum circuits and computing the expectation values of observables of interest. 

For the unraveling-based trajectory methods~\cite{PRXQuantum.4.040326,PhysRevLett.133.230403,cichy2025classicalsimulationnoisyquantum},
the mixed-state evolution generated by local decoherence channels is rewritten as an ensemble average over stochastic pure-state trajectories. In this framework, each noise channel is represented by a set of Kraus operators, and the corresponding pure-state evolution can be simulated efficiently using matrix-product-state methods provided that the entanglement growth along individual trajectories remains sufficiently weak. A key observation is that the Kraus representation of a quantum channel is not unique: different but equivalent unravelings are related by unitary rotations in the ancilla space and can lead to markedly different entanglement structures for the resulting trajectories. This freedom can therefore be exploited to optimize the unraveling and minimize the trajectory entanglement, thereby substantially enlarging the parameter regime in which noisy dynamics can be simulated efficiently. See Fig.~\ref{fig:unravelingFig1} for the illsutration of trajectory unraveling method.

For the Pauli-propagation or path-integral-inspired approaches~\cite{10.1145/3564246.3585234,PhysRevLett.133.120603,angrisani2025simulatingquantumcircuitsarbitrary,lee2025classicalsimulationnoisyrandom,j1gg-s6zb,GonzalezGarcia2025paulipath,li2025dualrolelowweightpauli,fontana2023classicalsimulationsnoisyvariational}, instead of evolving the full density matrix, this method propagates observables backward through the noisy circuit and expands the resulting operator in terms of Pauli paths. The key idea is that, under local incoherent noise, the contributions of long and high-weight Pauli paths are strongly suppressed, so that the expansion can be truncated efficiently. This leads to a practical algorithm for estimating expectation values in noisy circuits without explicitly tracking the full mixed-state evolution. Importantly, this framework applies to more general local incoherent noise channels, including dephasing and non-unital noise. Both analytical arguments and numerical simulations show that the method becomes increasingly effective as the noise strength grows, enabling the efficient simulation of large noisy quantum circuits in regimes where direct density-matrix methods are infeasible.

These algorithms, many of which leverage the structure imposed by noise, have broadened the toolkit for understanding quantum dynamics in practical, imperfect settings.

\subsection{Mixed-state phases of matter}

\begin{figure*}[t]
\centering
\includegraphics[width=0.5\textwidth, keepaspectratio]{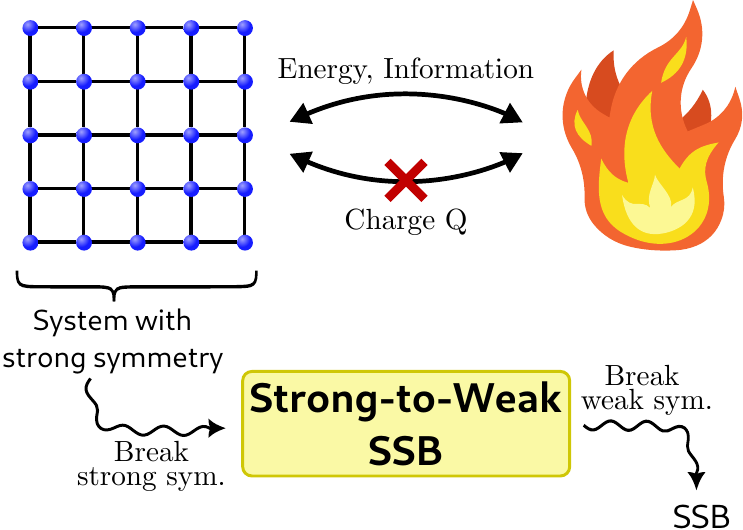}
\caption{\textit{Strong-to-weak spontaneous symmetry breaking}. For a mixed state with a strong symmetry coupled to an environment, strong-to-weak spontaneous symmetry breaking occurs when the strong symmetry is spontaneously broken while the corresponding weak symmetry remains unbroken. Reproduced from Ref.~\cite{PRXQuantum.6.010344}. \href{https://creativecommons.org/licenses/by/4.0/}{CC BY 4.0}.
}
\label{fig:SWSSB}
\end{figure*}

In the past few years, intensive progress has been made in extending the notion of phases of matter from pure ground states to mixed states~\cite{Coser2019classificationof,PhysRevLett.134.070403}, motivated by noisy devices in the NISQ era and the unavoidable coupling to the environment in realistic physical systems. 
Noisy quantum circuits naturally serve as a versatile platform for exploring mixed-state phases of matter.

Quantum phases at zero temperature can be characterized as equivalence classes under local unitary transformations: two ground states within the same gapped phase can be transformed into each other by a local unitary circuit, and the energy gap remains open along the interpolation between them. Analogously to the energy gap used to characterize pure-state quantum phases, the Markov length, defined as the characteristic length scale associated with the conditional mutual information, has been proposed as a criterion for classifying mixed-state phases of matter~\cite{PhysRevLett.134.070403}: the mixed state remains in the same phase as long as its Markov length stays finite throughout the evolution. Operationally, noisy circuit based characterizations of mixed states~\cite{PhysRevX.14.031044,PhysRevLett.134.070403,ma2025circuitbasedchatacterizationfinitetemperaturequantum,sang2025mixedstatephaseslocalreversibility} have also been proposed and applied to distinguish distinct mixed-state phases of matter.

Moreover, phase transitions in mixed states exhibit richer physics. For mixed states, there are two types of symmetries, namely strong symmetries and weak symmetries, whereas for pure states only strong symmetries arise. As a result, strong-to-weak spontaneous symmetry breaking has attracted increasing attention: for a mixed state with a strong symmetry coupled to an environment, strong-to-weak spontaneous symmetry breaking occurs when the strong symmetry is spontaneously broken while the corresponding weak symmetry remains unbroken~\cite{PRXQuantum.4.030317, gu2024spontaneoussymmetrybreakingopen,sala2024spontaneousstrongsymmetrybreaking, PhysRevB.111.064111, PRXQuantum.6.010344,guo2025quantumstrongtoweakspontaneoussymmetry, PRXQuantum.6.010314, PhysRevB.111.L201108, 7p5x-7yqb,liuDiagnosingStrongtoweakSymmetry2025, PhysRevB.111.L060304, ding2026strongtoweakspontaneoussymmetrybreaking}. A schematic illustration of strong-to-weak spontaneous symmetry breaking is shown in Fig.~\ref{fig:SWSSB}. Since pure states can only respect strong symmetries, there is no counterpart of strong-to-weak spontaneous symmetry breaking in pure-state phase transitions.
When the noise channel respects certain symmetries that are also respected by the quantum state, noisy quantum circuits provide a useful tool for investigating strong-to-weak spontaneous symmetry breaking both numerically and analytically, helping deepen our understanding of phases of matter.

\subsection{Error mitigation and error correction}
Furthermore, profound connections have been established between noisy monitored quantum circuits and quantum error correction.  
Measurement-induced phase transitions have been shown to generate emergent quantum error-correcting structures~\cite{PhysRevLett.125.030505, PhysRevB.111.064308, PhysRevX.10.041020}, linking dynamical entanglement transitions to threshold phenomena in fault-tolerant quantum computation.  
Related developments include advances in new dissipative or measurement-assisted mechanisms for constructing quantum error-correcting codes~\cite{bao2023mixedstatetopologicalordererrorfield, PRXQuantum.5.020343, hlfh-86yz, PhysRevA.111.032402}. 
Recently, topological order and symmetry-protected topological (SPT) phases have also been extensively investigated in noisy monitored quantum circuits~\cite{lavasaniMeasurementinducedTopologicalEntanglement2021d, PhysRevB.108.224304, yu2025gaplesssymmetryprotectedtopologicalstates,PhysRevB.111.064111, PhysRevLett.132.070401, PhysRevB.108.115135, PhysRevB.108.094304}.  
The topological order or SPT phases are robust against local (symmetric) perturbations, whose intrinsic topological degeneracy or edge modes provide precisely the kind of redundancy required for encoding quantum information.  
This perspective highlights how noisy monitored quantum circuits can act as a resource for building and stabilizing error-correcting code spaces.

\begin{figure*}[t]
\centering
\includegraphics[width=0.65\textwidth, keepaspectratio]{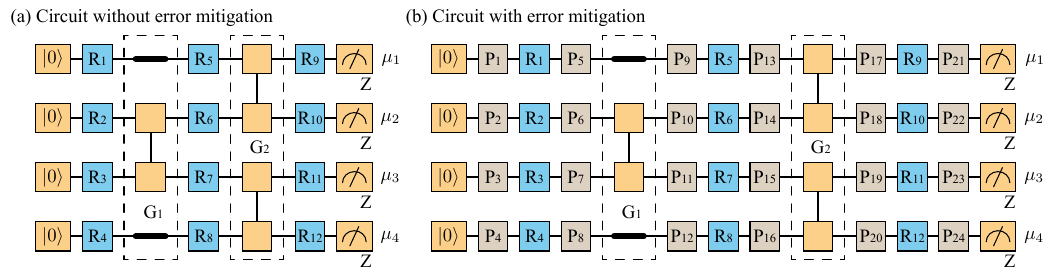}
\caption{\textit{Simple example of a circuit without and with error mitigation}. To implement the error mitigation, two layers of Pauli gates $P_i$
(brown) are introduced before and after each layer of computing gates (blue). Reproduced from Ref.~\cite{PRXQuantum.2.040330}. \href{https://creativecommons.org/licenses/by/4.0/}{CC BY 4.0}.
}
\label{fig:QEM}
\end{figure*}

On the other hand, quantum error mitigation~\cite{RevModPhys.95.045005} is necessary in order to extract reliable results or information from the output states of noisy quantum circuits and, more generally, noisy quantum devices. Various algorithms have been proposed, including zero-noise extrapolation~\cite{PhysRevX.7.021050,PhysRevLett.119.180509,kandalaErrorMitigationExtends2019,kimScalableErrorMitigation2023,tran2023localityerrormitigationquantum}, probabilistic error cancellation~\cite{PhysRevLett.119.180509,PhysRevX.8.031027,doi:10.1126/sciadv.aaw5686,qsmz-9kkh,tran2023localityerrormitigationquantum}, and virtual distillation~\cite{PhysRevX.11.041036,PhysRevLett.132.050203,Bultrini2023unifying}.
Recent work has established machine learning as a promising tool for quantum error mitigation in the NISQ regime~\cite{PRXQuantum.2.040330, bennewitzNeuralErrorMitigation2022, https://doi.org/10.1002/qute.202300147,liaoMachineLearningPractical2024, Kim_2022, nguyen2025diffusion, muqeet2024machinelearningbasederrormitigation,Czarnik2021errormitigation}. 
A central idea is to train a classical model to learn the map from noisy quantum outputs to their ideal counterparts, thereby reducing the sampling and runtime overheads of traditional mitigation protocols. Early learning-based QEM showed that one can train on classically tractable variants of a target circuit and then use the learned model to mitigate errors in more general circuits~\cite{PRXQuantum.2.040330} (see Fig.~\ref{fig:QEM}). More recently, machine-learning-based QEM has been demonstrated on quantum processors with up to $100$ qubits, using models such as linear regression, random forests, multilayer perceptrons, and graph neural networks, and was found to substantially reduce mitigation cost while maintaining competitive accuracy relative to standard zero-noise extrapolation~\cite{liaoMachineLearningPractical2024}. Deep-learning approaches have also been applied successfully to readout-error mitigation, where nonlinear correlations in measurement noise can be learned more effectively than with conventional linear-inversion methods~\cite{Kim_2022}. 

Beyond these supervised-learning protocols, recent studies have explored more flexible and task-adapted machine-learning strategies. In particular, diffusion-inspired frameworks have been proposed for mitigating noise in parameterized quantum circuits~\cite{nguyen2025diffusion}, while newer data-driven approaches aim to perform error mitigation without detailed prior knowledge of the noise model or even without noiseless training data. These developments suggest that machine learning can enhance the practicality and scalability of QEM, especially when the target task belongs to a restricted circuit family and sufficient training data are available.

In sum, these works demonstrate that noisy monitored quantum circuits not only illuminate the fundamental behavior of quantum systems under decoherence, but also offer concrete guiding principles for engineering noise-resilient quantum technologies, including pathways toward better quantum error correction and mitigation designs.

\section{Discussions and outlooks}

In this review, we have presented a unified perspective on noisy monitored quantum circuits and their role at the interface of quantum information, quantum many-body physics, and quantum computation.  
By mapping circuit dynamics to classical statistical models in one higher dimension, we surveyed how quantum noise fundamentally reshapes entanglement structure, eliminates the conventional measurement-induced phase transition, and gives rise to universal scaling behaviors such as the characteristic $q^{-1/3}$ entanglement scaling.  
We further reviewed the information dynamics and information protection capabilities of noisy circuits, emphasizing the distinct timescales emerging from temporally uncorrelated and temporally correlated noise, and clarified their connections to domain-wall fluctuations and the Hayden--Preskill protocol.  
In addition, we discussed the broader landscape of noise-induced phase transitions, including the noise-induced entanglement, coding, complexity phase transitions. 

Beyond these core results, we highlighted a wide array of applications inspired by noisy monitored circuits, ranging from new designs in variational quantum algorithms and new classical simulation techniques to mixed-state phases of matter and developments in quantum error mitigation and correction.  
These directions demonstrate that noisy circuits are not merely a theoretical abstraction, but a practical framework for understanding and harnessing the interplay between randomness, measurement, and decoherence.

Looking ahead, several promising avenues merit further exploration.  
First, extending the statistical-model framework to quantum noise with tunable correlation strength, and higher-dimensional architectures may reveal new universality classes and dynamical phases.  
Second, the information-theoretic viewpoint suggests deeper connections between noisy circuit dynamics, black-hole--inspired decoding protocols, and fault-tolerant quantum computation.  
Third, the growing experimental capabilities of superconducting qubits~\cite{kohMeasurementinducedEntanglementPhase2023,lavasaniMeasurementinducedTopologicalEntanglement2021d,hokeMeasurementinducedEntanglementTeleportation2023}, trapped ions~\cite{noelMeasurementinducedQuantumPhases2022b}, and Rydberg arrays~\cite{RevModPhys.82.2313,bluvsteinQuantumProcessorBased2022a,schollErasureConversionHighfidelity2023,doi:10.1126/science.ade5337,bluvsteinLogicalQuantumProcessor2024a,xuConstantoverheadFaulttolerantQuantum2024,bernien_probing_2017,scholl_quantum_2021,ebadi_quantum_2021,doi:10.1126/science.abg2530,6722-tf9c, Wu_2021, doi:10.1126/science.abi8794,PhysRevLett.133.223401} provide an exciting opportunity to probe noise-induced phenomena in regimes inaccessible to classical simulation.  
Finally, the interplay between noise and structure, such as symmetry, topology, long-range interactions, or tailored dissipative mechanisms, may enable the design of new dynamical phases, noise-resilient algorithms, and emergent error-correcting patterns.

Taken together, these developments position noisy monitored quantum circuits as a powerful and versatile platform for uncovering universal features of quantum dynamics in realistic, decohering environments, and for guiding the next generation of quantum technologies.

\section{Acknowledgement}
SXZ acknowledges the support
from Innovation Program for Quantum Science and
Technology (2024ZD0301700) and the National Natural Science Foundation of China (No. 12574546). SL's work at Princeton University was supported by the Gordon and Betty Moore Foundation
through Grant No. GBMF8685 toward the Princeton theory
program, the Gordon and Betty Moore Foundation’s EPiQS
Initiative (Grant No. GBMF11070), the Global
Collaborative Network Grant at Princeton University,
the Simons Investigator Grant No. 404513, the NSF-MERSEC (Grant No. MERSEC DMR 2011750), the Simons
Collaboration on New Frontiers in Superconductivity,
and the Schmidt Foundation at the Princeton University. The work of S.-
K.J. is supported by a start-up fund at Tulane University.

\bibliographystyle{apsreve}
\bibliography{ref}
\end{document}